\documentclass[a4paper,11pt]{article}
\pdfoutput=1 % if your are submitting a pdflatex (i.e. if you have
\usepackage{jcappub} % for details on the use of the package, please
\usepackage[T1]{fontenc} % if needed
\usepackage{mathtools}

\usepackage{enumitem}

\subheader{\footnotesize
        \vspace*{-3.7em}
        \begin{flushright}
        \end{flushright}
}

\usepackage{aas_macros}
\usepackage{amsmath,amssymb}
\usepackage{cleveref}
\usepackage{siunitx}
\usepackage{enumerate}
\usepackage[dvipsnames]{xcolor}

\usepackage{tikz-feynman}
\usepackage{caption}
\usepackage{subcaption}

\usepackage{amsthm}
\usepackage{makecell}

\usepackage{psfrag}
\usepackage{color}
\usepackage{graphicx}
\usepackage{upgreek}
\usepackage{feynmp}
\DeclareMathAlphabet{\mathpzc}{OT1}{pzc}{m}{it}

\usepackage{tikz}
\usetikzlibrary{trees}
\usetikzlibrary{decorations.pathmorphing}
\usetikzlibrary{decorations.markings}
\tikzset{
    photon/.style={decorate, line width=0.15mm, decoration={snake,amplitude=3pt,segment length=8pt}, draw=black},
    wino/.style={draw=redwine},    
    fermion/.style={draw=black, line width=0.2mm, postaction={decorate},
        decoration={markings,mark=at position .55 with {\arrow[draw=black,scale=2,#1]{>}}}},
    scalar/.style={draw=black, dashed,postaction={decorate},
        decoration={markings,mark=at position .55 with {\arrow[draw=black,scale=2,#1]{>}}}},
    scalarline/.style={draw=black, postaction={decorate},
        decoration={markings,mark=at position .55 with {\arrow[draw=black,scale=2,#1]{>}}}},
    scalarline2/.style={draw=black, postaction={decorate} },
    scalar2/.style={draw=black, dashed,postaction={decorate}},
    gluon/.style={decorate, draw=black,
        decoration={coil,amplitude=3pt, segment length=4pt}},
    graviton/.style={decorate, draw=black,
        decoration={zigzag,amplitude=3pt, segment length=4pt}}
}
\tikzstyle{blob}=[circle,
                              minimum size=20,
                              draw=black!80,
                              fill=black!80]   
\tikzstyle{redblob}=[circle,
                              thick,
                              minimum size=0.4cm,
                              draw=red!80,
                              fill=red!60]

\allowdisplaybreaks[4]

\definecolor{darkgreen}{rgb}{0,0.5,0}
\newcommand{\beq}{\begin{eqnarray}}
\newcommand{\eeq}{\end{eqnarray}}

\newcommand{\comment}[1]{}

\newcommand{\dt}{{\rm d}t}
\newcommand{\Nres}{{\cal N}}
\newcommand{\Mest}{M_{s,{\rm est}}}
\newcommand{\rhop}{\rho_{\rm max}}

\newcommand{\tend}{t_{\rm end}}

\newcommand{\rhoc}{\rho_{s,\,{\rm peak}}}

\newcommand{\rinit}{r_{s}^{\rm init}}

\newcommand{\bseq}{\begin{subequations}}
\newcommand{\eseq}{\end{subequations}}
\newcommand{\be}{\begin{equation}}
\newcommand{\ee}{\end{equation}}

\def\d{\partial}
\def\e{{\rm e}}
\def\d{\partial}

\renewcommand{\ln}{\mathop{\rm ln}\nolimits}

\renewcommand{\Im}{\mathop{\rm Im}\nolimits}

\newcommand{\ve}{\varepsilon}

\renewcommand{\k}{\mathbf{k}}
\newcommand{\x}{\mathbf{x}}
\newcommand{\E}{\mathcal{E}}

\newcommand{\beqa}{\begin{eqnarray}}
\newcommand{\eeqa}{\end{eqnarray}}

\newcommand\m{\mu}

\newcommand\vf{\varphi}

\title{
Condensation and Evaporation of Boson Stars} 

\author[a,b,c]{James Hung-Hsu Chan,}
\author[d,e]{Sergey~Sibiryakov,}
\author[f]{and Wei~Xue}

\affiliation[a]{
Department of Astrophysics, American Museum of Natural History,\\
Central Park West and 79th Street, NY 10024-5192, USA} 
\affiliation[b]{
Department of Physics and Astronomy, Lehman College of the CUNY,\\
Bronx, NY 10468, USA }
\affiliation[c]{
Institute of Physics, Laboratory of Astrophysique,
\' Ecole Polytechnique F\' ed\' erale\\ de Lausanne (EPFL),
Observatoire de Sauverny, 1290 Versoix, Switzerland}
\affiliation[d]{Department of Physics \& Astronomy, McMaster
University,\\
Hamilton, Ontario, L8S 4M1, Canada} 
\affiliation[e]{Perimeter Institute for Theoretical Physics, Waterloo,
 Ontario, N2L 2Y5, Canada}
\affiliation[f]{Department of Physics, University of Florida,
  Gainesville, FL 32611, USA} 

\emailAdd{jchan@amnh.org}
\emailAdd{ssibiryakov@perimeterinstitute.ca}
\emailAdd{weixue@ufl.edu}

\abstract{
Axion-like particles, including the QCD axion, are well-motivated dark matter candidates. Numerical simulations have revealed coherent soliton configurations, also known as boson stars, in the centers of axion halos. We study evolution of axion solitons immersed into a gas of axion waves with Maxwellian velocity distribution. Combining analytical approach with controlled numerical simulations we find that heavy solitons grow by condensation of axions from the gas, while light solitons evaporate. We deduce the parametric dependence of the soliton growth/evaporation rate and show that it is proportional to the rate of the kinetic relaxation in the gas. The proportionality coefficient is  controlled by the product of the soliton radius and the typical gas momentum or, equivalently, the ratio of the gas and soliton virial temperatures. We discuss the asymptotics of the rate when this parameter is large or small.
}

\begin{document}
\maketitle
\flushbottom

%==========================================================================
\section{Introduction}
\label{sec:intro}
%==========================================================================

QCD axion
\cite{Weinberg:1977ma,Wilczek:1977pj,Shifman:1979if,Kim:1979if,Zhitnitsky:1980tq,Dine:1981rt,Preskill:1982cy,Abbott:1982af,Dine:1982ah} 
and axion-like particles
\cite{Arvanitaki:2009fg,Marsh:2015xka,Hui:2016ltb,Hui:2021tkt}, are widely
discussed in the literature as well-motivated dark matter (DM)
candidates. 
The QCD axion, originally suggested as a solution to the strong CP
problem \cite{Peccei:1977hh,Peccei:1977ur}, was soon realized
\cite{Preskill:1982cy} to be produced in the early universe and
behave as cold 
dark matter after the QCD phase transition endowing it with the
mass. The requirement that the QCD axion accounts for all of DM leads
to a preferred mass window\footnote{The mass can be
  smaller in scenarios where Peccei--Quinn symmetry is never restored
  after inflation.}
 $m_{aQCD}\sim 10^{-6}\div 10^{-4}~{\rm eV}$. 

Axion-like particles with broad range of masses and very weak coupling
to 
the
Standard Model naturally arise in many beyond Standard Model
scenarios and string theory
\cite{Svrcek:2006yi,Arvanitaki:2009fg}. For brevity, we will refer to
DM made of such particles as axion DM.  
Particularly interesting is the case of ultralight (also called ``fuzzy'')
DM with mass $m_a\sim 10^{-22}\div 10^{-19}~{\rm eV}$
\cite{Hu:2000ke}. The de Broglie wavelength of such ultralight
particle corresponding to virial velocity in a galactic 
halo,\footnote{Throughout the paper we
  use the system of units $\hbar=c=1$.}   
\begin{equation}
   \lambda_a   = \frac{2\pi }{ m v_a} \sim \, 1.2\times
      \, \left( \frac { m_a}{10^{-22}  \, \mathrm{eV} } \right)^{-1}
      \,  \left(  \frac{ v_a}{100\,\mathrm{km/s}} \right)^{-1}   \mathrm{kpc} \ ,
\end{equation}
is comparable to the typical cosmological and astrophysical
distances. Due to this property, ultralight dark matter exhibits rich
phenomenology affecting various cosmological observables and galactic
dynamics \cite{Marsh:2015xka,Hui:2016ltb,Hui:2021tkt}. The analysis of
Lyman-$\alpha$ forest
\cite{Armengaud:2017nkf,Kobayashi:2017jcf,Rogers:2020ltq}, galactic
rotation curves \cite{Bar:2018acw,Bar:2021kti}, halo
profiles of dwarf galaxies
\cite{Safarzadeh:2019sre,Zoutendijk:2021kee} and subhalo population in
the Milky Way \cite{DES:2020fxi} strongly disfavor DM lighter than
$10^{-21}~{\rm eV}$. Dynamical heating of stars by 
ultralight DM in ultrafaint dwarf
galaxies has been used to infer tighter constraints 
${m_a\gtrsim 10^{-19}~{\rm eV}}$ \cite{Marsh:2018zyw,Dalal:2022rmp}.

A distinctive feature of axion DM is its 
huge occupation numbers (phase-space density)
which are allowed because axions are bosons,
\begin{equation}
   f_{\textbf{k}}  \sim 10^{86}  \times  \left( \frac{ \rho_a} { 0.3
       \, {\rm GeV/cm}^3 }\right)
      \, \left( \frac{m_a} {10^{-20}  \, \mathrm{eV} }  \right)^{-4}
      \,  \left( \frac{v_a}{100\, {\rm km/s}} \right)^{-3}   \ .
   \label{eq:Na}
\end{equation}
This implies that, rather than behaving as a collection of individual
particles, axion DM is best described by a coherent classical scalar
field with the 
scattering rate of axions
increased due to the Bose enhancement. Typically, 
in the study of structure
formation 
all axion
interactions besides gravity can be neglected resulting in a
universal wave dynamics described by 
Schr\"odinger--Poisson equations \cite{Hui:2021tkt}. The dependence of
these equations on the axion mass can be taken into account by a simple
rescaling, and thus they apply to any axion DM 
as long as $f_{\bf k}\gg 1$.

The Schr\"odinger--Poisson system admits a spherically symmetric
localized solution known as {\it axion soliton} or {\it boson
  star}\footnote{We will use the two names 
  interchangeably.} 
\cite{Ruffini:1969qy}.
All axions comprising the soliton
are
in the same state which is the 
ground state of the gravitational potential and hence the soliton 
can be viewed
as inhomogeneous 
Bose--Einstein condensate sustained by its own gravity
 \cite{Guzman:2006yc}. 
Numerical simulations of axion DM have revealed formation of boson
stars in the centers of virialized
axion halos (also known as miniclusters
\cite{Hogan:1988mp,Kolb:1993zz} 
in the
case of QCD axion). This phenomenon was observed in the cosmological
setting \cite{Schive:2014dra,Veltmaat:2018,Mina:2020eik,May:2021wwp},
in numerical
experiments with halos created by collisions of several seed solitons
\cite{Schive:2014hza,Schwabe:2016rze,Mocz:2017wlg},
and in the kinetic relaxation regime~\cite{Levkov:2018kau}. 
It was also found that 
if the soliton is artificially
removed from the halo, evolution readily reinstates
it back~\cite{Yavetz:2021pbc}. 

Thus, presence of a solitonic core appears to be a generic feature of an
axion halo. The rest of the halo represents a cloud of 
randomly moving wavepackets with the velocities
roughly following the Maxwellian distribution and the average density
fitted by the NFW profile \cite{Navarro:1996gj}, 
similarly to the usual
cold DM. 
It is natural to ask how the soliton interacts with this environment. 
Refs.~\cite{Eggemeier:2019jsu,Schive:2019rrw,Li:2020ryg,Zagorac:2021}
showed that interference between the soliton and wavepackets leads to
oscillations of its density and to a random walk of the soliton center
around the halo center of mass. Further, 
an interesting correlation between the soliton mass and the mass of
its host halo has been established in cosmological numerical
simulations \cite{Schive:2014dra,Schive:2014hza} and confirmed in
\cite{Veltmaat:2018,Eggemeier:2019jsu}. This relation can be rephrased
as equality between the virial temperatures of the soliton and the
host halo. While this relation may appear intuitive, the physical
mechanism behind it remains unclear. It is not reproduced by 
simulations starting from
non-cosmological initial conditions
\cite{Schwabe:2016rze,Mocz:2017wlg,Chan:2021bja}, whereas more recent
cosmological simulations \cite{Nori:2020jzx,
May:2021wwp,Chan:2021bja} indicate that
it is subject to a large scatter, perhaps due to different merger
histories of different halos. The results of
Ref.~\cite{Levkov:2018kau} disfavor a potential interpretation of
the soliton-host halo relation as a condition for kinetic
equilibrium. Indeed, it was observed that, once formed, the solitons 
continue to grow by condensation of axions from the surrounding
gas. On the other hand, Refs.~\cite{Eggemeier:2019jsu,Chen:2020cef}
argue that this growth slows down when the soliton
becomes heavy enough to heat up the inner part of the halo and,
given the finite time of the simulations, this can explain the observed
correlation. The mass of the soliton can be also significantly
affected by baryonic matter, typically leading to its
increase \cite{ChanEtal18,Veltmaat:2019hou}. 

Boson stars give rise to
important signatures opening up various opportunities for future
discovery or constraints on axion DM. 
In the case of fuzzy DM, they are expected to
play a prominent role in galactic dynamics modifying the rotation
curves \cite{Bar:2018acw,Bar:2021kti} and heating the
stars in the central regions through oscillations and random walk
\cite{Marsh:2018zyw,Chiang:2021uvt,Chowdhury:2021zik}.
When axion self-interaction is included, they become unstable if
their mass exceeds a certain threshold and collapse producing bursts
of relativistic axions \cite{Levkov:2016rkk}. Further allowing for
possible axion coupling to photons, they can be sources of radio
emission \cite{Tkachev:2014dpa,Hertzberg:2018zte,Levkov:2020txo}.  
Presence or absence of boson stars in axion miniclusters can have
important implications for their density profiles and 
lensing searches \cite{Kavanagh:2020gcy,Ellis:2022grh}.
Very dense boson stars made of inflaton field get produced in
inflationary models with delayed reheating opening a potentially rich
phenomenology, such as seeding primordial
black holes or contributing into stochastic high-frequency
gravitational wave background \cite{Eggemeier:2021smj}.

The dynamical range achievable in axion DM simulations is severely
limited by the computational costs (see the discussion in
\cite{May:2021wwp}). This calls for better theoretical understanding
of the physical laws governing the evolution of boson stars in various
environments which would allow their extrapolation outside of the
parameter regions explored in simulations. In the present 
paper we make a step
in this direction by studying the evolution of a boson star immersed
in a box filled with homogeneous axion gas. Focusing on this setup
allows us to get rid of the uncertainties related to the dynamics of
the halo and keep under control the gas density and its velocity
distribution. The latter is chosen to be Maxwellian at the initial
moment of time. Similar setup was
employed in Ref.~\cite{Levkov:2018kau} to study the formation of the
soliton in the process of the gas kinetic relaxation. By contrast, we do not
assume the soliton to be formed from the gas and simply add it in the
initial conditions of our simulations. In this way we are able to
explore a wide range of soliton masses corresponding different ratios
between the soliton virial temperature $T_s$ and the temperature of
the gas $T_g$.

The key quantity that we are interested in is the rate of change of
the soliton mass,
\begin{equation}
\label{rate_def}
\Gamma_s=\frac{1}{M_s}\frac{dM_s}{dt}\;.
\end{equation}
We study the dependence of this quantity on the parameters
characterizing the gas and the soliton by a combination of analytical
and numerical methods.
We find that the solitons with $T_s/T_g\gtrsim 0.1$ grow by absorbing
particles from the gas. For fixed gas parameters, the growth rate is
essentially constant in the range $0.1\lesssim T_s/T_g\lesssim 1$,
whereas at $T_s/T_g\gtrsim 1$ it decreases as $(T_s/T_g)^{-n/2}$
with $n=2\div 4$. 

Interestingly, we find that if $T_s/T_g\lesssim 0.08$, the soliton
{\em evaporates}, the time scale of this process being parametrically
shorter than the relaxation time. 
This does not contradict previous results on soliton
formation from the gas by kinetic relaxation
\cite{Levkov:2018kau}.
Indeed, by running the simulations
longer than the evaporation of the initial soliton we observe after a while
the birth of a
new soliton with $T_s/T_g\gtrsim 0.1$, 
in agreement with \cite{Levkov:2018kau}. It is
worth stressing the difference between the soliton evaporation and tidal
disruption by large-scale gradients of the halo gravitational field
\cite{Du:2018qor}. This is clear already from the fact that there is no
halo in 
our 
setup. Moreover, the qualitative direction of the process ---
evaporation vs. condensation --- 
is
entirely determined by the soliton and gas temperatures and does not
depend on the density contrast between them.\footnote{Though the
  quantitative characteristics --- the evaporation rate --- does
  depend on the gas density,  
 $\Gamma_s\propto
  \rho_g^2$ (see eq.~(\ref{Gammagamma})).}
 
The paper is organized as follows. In \cref{sec:Maxwell} we
introduce our framework and review the relevant properties of the
soliton solution to the Schr\"odinger--Poisson equations. In 
\cref{sec:theory} we address the computation of the soliton
growth/evaporation rate formulating it as a quantum-mechanical 
scattering
problem. We consider separately the cases of light (cold, $T_s/T_g\ll 1$)
and heavy (hot, $T_s/T_g\gg 1$) solitons and employ various approximations
to estimate the rate analytically. In 
\cref{sec:simulation} we describe our numerical simulations,
extract the soliton growth rate from them 
and compare it to the analytic predictions.  
In \cref{sec:conclusion} we discuss the implications of our results
and compare to other works. Three appendices contain auxiliary
material. 
In \cref{app:class} we provide an alternative derivation of the
soliton growth rate using only classical equations of motion. In
\cref{sec:levkov} we describe a suit of simulations
reproducing the setup of Ref.~\cite{Levkov:2018kau} where the soliton
forms from the gas spontaneously due to kinetic
relaxation. Appendix~\ref{sec:nums} contains additional details about
our numerical procedure.

\section{Soliton Wavefunction and Axion Gas}
\label{sec:Maxwell}

Non-relativistic axions with mass $m$ are described by a complex
scalar field $\psi$ obeying the Schr\"odinger--Poisson equations, 
\bseq
\label{eq:SPeq}
\begin{align}
&i\d_t\psi +\frac{\Delta\psi}{2m}-m\Phi\psi=0\;,\\
&\Delta\Phi=4\pi G m\,|\psi|^2\;,
\end{align}
\eseq
where $G$ is the gravitational coupling, $\Phi$ is the Newton
potential and 
$\Delta$ denotes the Laplacian.
The square of the field gives the particle number density,
$|\psi(t,\x)|^2=n(t,\x)$. Equations (\ref{eq:SPeq}) are invariant
under scaling transformations,
\bseq
\label{eq:scaling}
\begin{gather}
     \psi \mapsto \tilde\psi(t,\x)=\Lambda_3 
\psi  (\Lambda_1 t ,\Lambda_2\x)  \, ,
     \qquad
     \Phi\mapsto \tilde\Phi( t, \x ) =
\frac{\Lambda_1^2}{\Lambda_2^2} \Phi (\Lambda_1t,\Lambda_2 \x) \, ,\\
m\mapsto \tilde m=\frac{\Lambda_2^2}{\Lambda_1}m\,,\qquad
G\mapsto \tilde G=\frac{\Lambda_1^3}{\Lambda_2^2\Lambda_3^2}G\,,
\end{gather}
\eseq
where $\Lambda_{1,2,3}$ are arbitrary parameters. A one-parameter
family of these transformations that leaves $m$ and $G$ invariant
connects different solutions for a given axion; the
transformations that change the mass, but not $G$, allow one to map
between 
solutions for axions with different masses; finally, the rescaling of
$G$ provides a freedom in the choice of units which is handy in
numerical simulations.

\begin{figure}[t]
\begin{center}
 \includegraphics[width=0.45\textwidth]{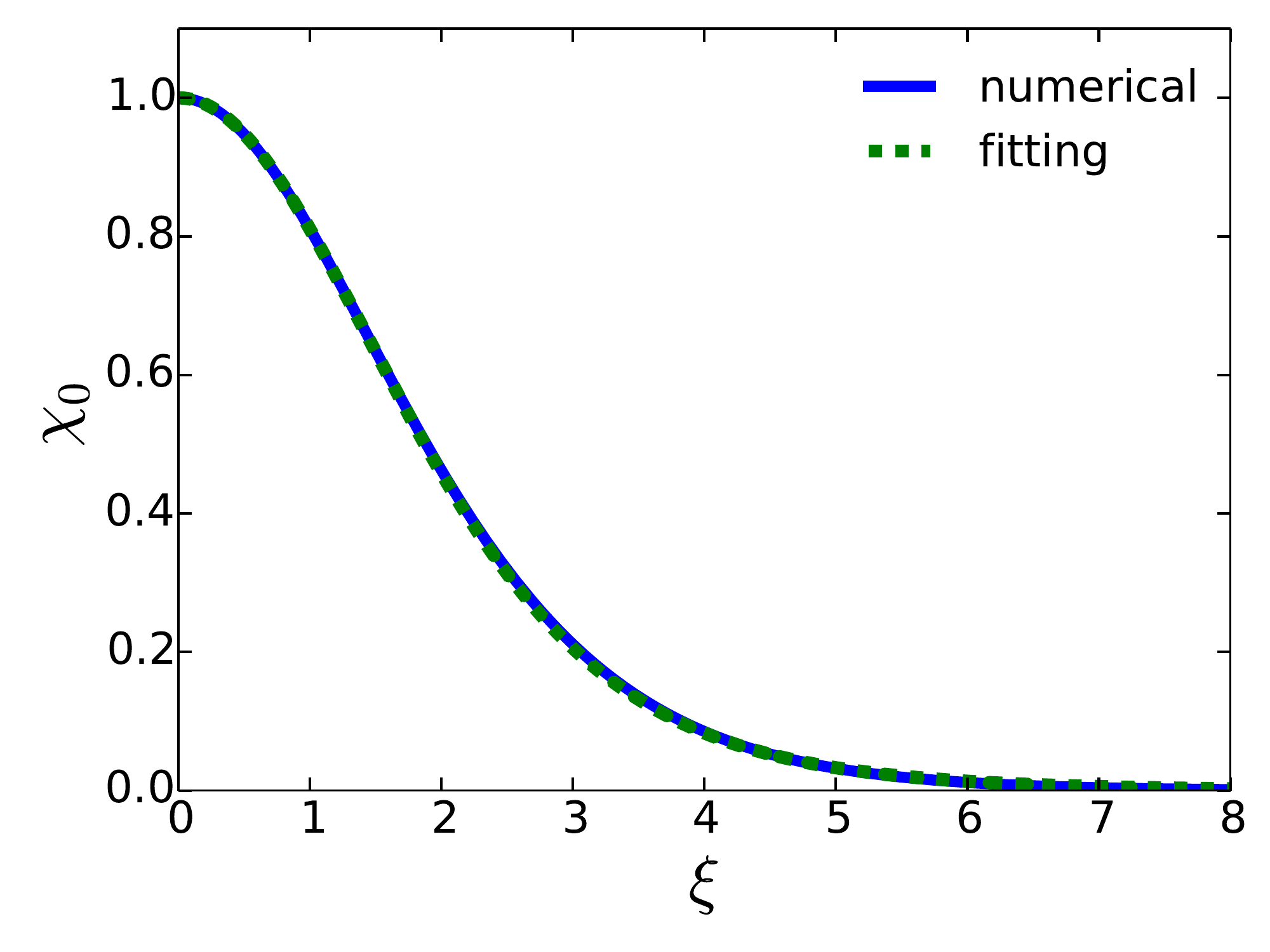}
\qquad
\quad
 \includegraphics[width=0.45\textwidth]{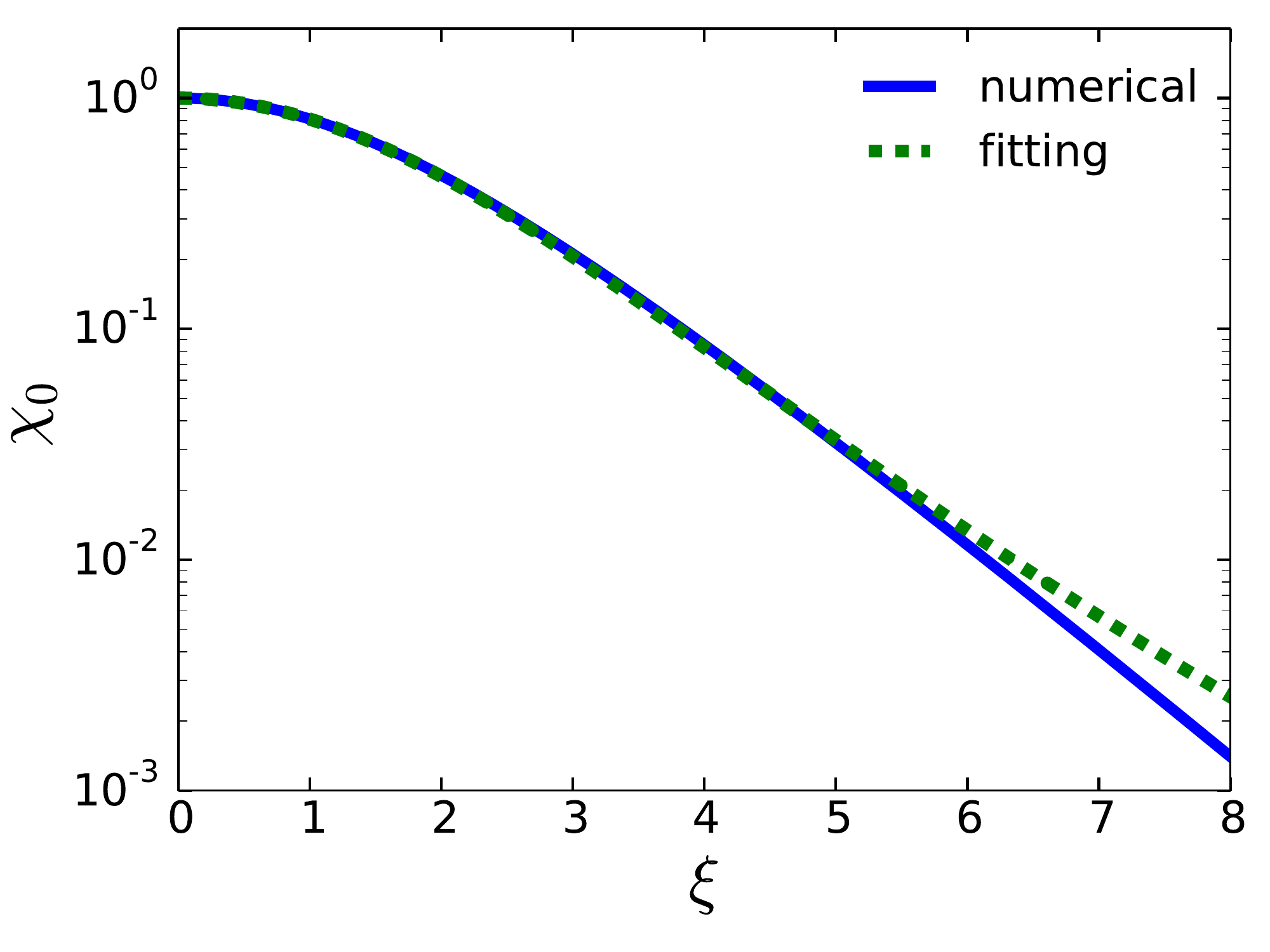}
\caption{The standard soliton profile in linear (left) 
and in log (right) scale.
The solid lines show the exact solution of the Schr\"odinger--Poisson
equations,
while the dotted lines 
correspond to the fitting function (\ref{chifit}).    
  \label{fig:standsol}
}
\end{center}
\end{figure}

The system (\ref{eq:SPeq}) admits periodic spherically symmetric
solutions of the form,
\be
\label{soliton}
\psi_s(t,\x)=\chi(|\x|) \e^{ - i {\cal E}_s t}\;.
\ee 
The corresponding density $\rho_s(\x)=m|\chi(|\x|)|^2$ is
time-independent and localized in space, hence these solutions are
called {\rm solitons}. ${\cal E}_s$ represents the binding energy
(chemical potential) of axions in the soliton and is negative. There
is a continuous family of solitons differing by their mass $M_s$ and
related by the subgroup of the scaling transformations
(\ref{eq:scaling}) that leave $m$ and $G$ fixed. Using this symmetry,
the soliton wavefunction can be written as
\be
\label{solitonWF}
\chi(x)=\frac{k_s^2}{\sqrt{4\pi G m^3}}\chi_0(k_s x)\;,
\ee
where $k_s$ is the scaling parameter characterizing the soliton
width. By the uncertainty relation, it sets the typical momentum of
particles comprising the soliton. The dimensionless function
$\chi_0(\xi)$ describes the ``standard soliton'' normalized by the
condition 
\bseq
\label{stndeq}
\be
\label{standbc}
\chi_0(0)=1\;.
\ee
It solves the eigenvalue problem following from the
Schr\"odinger--Poisson system,
\begin{align}
\label{standeq1}
&\chi_0''+\frac{2}{\xi}\chi_0'=2(\Phi_0-\varepsilon_0)\chi_0\;,\\
\label{standeq2}
&\Phi_0''+\frac{2}{\xi}\Phi_0'=\chi_0^2\;,
\end{align}
\eseq
where $\Phi_0(\xi)$ is the standard soliton gravitational potential
and $\ve_0$ is its binding energy. Fig.~\ref{fig:standsol} shows the
function $\chi_0(\xi)$ obtained by numerically solving
eqs.~(\ref{stndeq}). It is well approximated by an analytic fit,
\be
\label{chifit}
\chi_{0,{\rm fit}}=\big(1+c_0\xi^2\big)^{-4}~,~~~~~~c_0=0.0539\;,
\ee
also shown in the figure. The fit differs from the exact solution only
at the tail where the exact solution falls off exponentially, whereas
the fit behaves as a power-law.

The standard soliton is characterized by the following dimensionless
quantities: 
\bseq
\label{standnums}
\begin{align}
&\ve_0=-0.692&& \text{binding energy}\;,\\
&\mu_0=4\pi \int_0^\infty d\xi\, \xi^2
\chi_0^2(\xi)  = 25.9&& \text{total mass}\;,\\
&\xi_0=1.299&&\text{half-density radius,}~|\chi_0(\xi_0)|^2=1/2\;.
\end{align}
\eseq 
The corresponding values for a general soliton are obtained by
rescaling,
\be
\label{EsMsrs}
{\cal E}_s=\ve_0\frac{k_s^2}{m}~,~~~~~~
M_s=\mu_0 \frac{k_s}{4\pi Gm^2}~,~~~~~~
r_s=\frac{\xi_0}{k_s}\;,
\ee
and its density profile can be approximated as
\begin{equation}
   \rho_s ( {\bf x}  )  
    \approx \frac{\rho_{s,\,{\rm peak}}}{  
\left[ 1  + c_s \, \left( |{\bf x}|/ r_{s}  \right)^2 \right]^8
}~,~~~~~
\rho_{s,\,{\rm peak}}=\frac{k_s^4}{4\pi G m^2}~,~~c_s=0.091\;.
   \label{eq:rhos}
\end{equation}
Note that the width of the soliton is inversely proportional to its
mass. Accordingly, the peak density is proportional to the fourth
power of the mass. The total energy of the soliton consists of kinetic
and potential parts,
\be
\label{Esoltot}
E_s=E_{s,{\rm kin}}+E_{s,{\rm pot}}=\int d^3x
\left(\frac{|\nabla\psi_s|^2}{2m} +\frac{m\Phi_s|\psi_s|^2}{2}\right)\;.
\ee
Using the Schr\"odinger--Poisson equations one can show that they obey
the virial theorem, $E_s=-E_{s,{\rm kin}}=E_{s,{\rm pot}}/2$, and
\be
E_s=\frac{M_s \E_s}{3m}\;.
\ee
It is instructive to introduce the soliton virial temperature,
\be
\label{Ts1}
T_s=\frac{2m E_{s,{\rm kin}}}{3M_s}=-\frac{2}{9}\E_s\;.
\ee
Using eqs.~(\ref{EsMsrs}) one obtains alternative expressions,
\be
\label{Ts2}
T_s=0.154\frac{k_s^2}{m}=\frac{0.259}{mr_s^2}\;.
\ee

We are interested to study how the mass of the soliton
varies due to its interaction with a gas of axion waves.
We assume the gas to fill
a box
of size 
\be
\label{Lrs}
L\gg r_s\;.
\ee 
Far away from the soliton, 
it is
described by a collection of plane waves,\footnote{At $|\x|\lesssim
  r_s$ the wavefunctions are modified by the gravitational field of
  the soliton, see below.}
\be
\label{psigas}
\psi_g(t,\x)=\frac{1}{L^{3/2}}\sum_\k a_{\k}\,
\e^{-i\frac{k^2}{2m}t+i\k\x}~,~~~~~~~|\x|\gg r_s\;.
\ee
We choose the occupation numbers to follow the Maxwell distribution,
consistent with the velocity distribution in a DM halo,
\be
\label{fgas}
f_\k\equiv |a_{\k}|^2 =  f_g\,\e^{-k^2/k_g^2}\;,
\ee 
where $k_g$ sets the characteristic momentum of particles in the
gas. The normalization $f_g$ is related to the gas density as
\be
\label{fgrhog}
f_g=\frac{(4\pi)^{3/2}}{m k_g^3}\rho_g\;.
\ee
Validity of the classical description requires $f_g\gg 1$.
The phases of the amplitudes $a_{\k}$ are assumed to be random. 

Using $k_g$ we can define an effective gas temperature,
\be
\label{Tgas}
T_g=\frac{k_g^2}{2m}\;.
\ee
To avoid confusion, we stress that this is not a true thermodynamic
temperature since \cref{fgas} is not an equilibrium distribution of
the boson gas which should follow the
Bose--Einstein formula. However, the latter cannot be reached within the
classical field theory. Rather, as demonstrated in
Ref.~\cite{Levkov:2018kau}, 
a homogeneous axion gas with initial distribution (\ref{fgas})
will evolve
towards the Rayleigh--Jeans occupation numbers diverging at low
$k$. This relaxation proceeds on the time scale
\be
\label{trelax}
\tau_{\rm rel}=
\frac{\sqrt{2}b\,k_g^6}{12\pi^3G^2m^3\rho_g^2\,
  \ln(k_gL)}\;,~~~~b\approx 0.9\;,
\ee
and culminates in the spontaneous formation of a soliton. 
We neglect the change of the gas distribution
in our theoretical considerations and discuss the validity of
this simplification later on. Numerically, we observe
that the Maxwell distribution appears to get reinstated in the gas once the
soliton is formed. Moreover, in the simulations where the soliton is
present for the whole duration, the distribution remains close to
Maxwellian at all moments of time.

Being a self-gravitating system, the homogeneous axion gas is unstable
with respect to gravitational collapse leading to a halo
formation. 
The corresponding Jeans length
is 
\be
\label{lJeans}
l_J=\frac{k_g}{m}\sqrt{\frac{\pi}{2G\rho_g}}\;,
\ee
where we have used that the sound speed in non-relativistic
Maxwellian gas is $k_g/(\sqrt{2}m)$. We avoid this instability by
considering the box size smaller than the Jeans length,
\be
\label{LlJ}
L<l_J\;.
\ee 
Note that this condition is compatible with \cref{Lrs} since $l_J$ can
be made arbitrarily large by decreasing the gas density. In practice,
however, \cref{LlJ} imposes strong limitations on the numerical
simulations, see \cref{sec:simulation}.

The total axion field describing a soliton immersed in the gas is
given by the sum 
\be
\label{split}
\psi(t,\x)=\psi_s(t,\x) + \psi_g(t,\x)\;.
\ee
For this decomposition 
to be well-defined, the number of particles in the
soliton must be much larger than in any other state in the gas,
\be
\label{largeNs}
M_s/m\gg f_\k\;.
\ee
To compare the soliton size with the characteristic wavelength of
axion waves, we introduce
\begin{equation}
   \nu \equiv \frac{k_g}{k_s} = 0.773\,r_s k_g
=0.555\sqrt{\frac{T_g}{T_s}}  \ .
   \label{eq:alpharatio}
\end{equation}
Recalling that the mass of the soliton is inversely proportional to
its size, we split solitons into three groups: 
{\it light solitons} ($\nu \gg 1$), {\it heavy solitons} ($\nu \ll
1$), and {\it median solitons} ($\nu \sim 1$). Note that light
solitons are also {\it cold}, heavy solitons are {\it hot}, whereas
median solitons have the same virial temperature as the gas. We are
going to see that the evolution of solitons from different groups is
dramatically different.

\section{Particle Exchange between Soliton and Gas}
\label{sec:theory}

\subsection{Soliton growth rate from wave scattering}
\label{sec:solitonrate}

Soliton is composed of Bose--Einstein condensate occupying the ground
state in its own gravitational potential. Several processes affect the
soliton in the axion gas. One of them is the interference of gas waves
with the soliton field which leads to fluctuations of its peak
density. Another one is elastic scattering of waves on the soliton
which endows it with momentum and leads to its Brownian motion. These
processes, however, do not change the number of particles in the
ground state and are not of interest to us. We focus on the processes
that lead to particle exchange between the gas and the soliton and
thereby affect the amplitude of the Bose--Einstein condensate.
In this section we develop their description
using scattering
theory. We adopt the language of quantum field theory as the most
convenient tool for this task. However, it is important to emphasize
that quantum physics is not essential for the soliton-gas
interaction. In \cref{app:class} we show how the same results can be
obtained within purely classical approach.

We start by observing that the Schr\"odinger--Poisson equations can be
derived from the action
\be
\label{Saxion}
S=\int dtd^3x\,\bigg(i\psi^*\d_t\psi+\frac{\psi^*\Delta\psi}{2m}
+\frac{\Phi\Delta\Phi}{8\pi G}-m\Phi |\psi|^2\bigg)\;.
\ee
We decompose the total axion field into the soliton and gas
components as in \cref{split}. At this point we should be more precise
about how we perform the split. The spectrum of particle states in the
soliton background contains unbound states with wavefunctions
becoming plane waves far away from the soliton, as well as bound
states in the soliton gravitational potential. In the literature, 
the latter are usually
interpreted as excitations of the soliton. While
this is a valid interpretation, it is more convenient
for our purposes 
to include them into the gas. The physical reason 
is that no matter whether the state is bound or not, a
transfer of particles to it from the ground state will deplete
the coherence of the soliton, whereas the inverse process clearly has
an opposite effect. Thus, we adopt the following
convention: the soliton component refers to coherent particles
strictly in the ground state described by the wavefunction
(\ref{soliton}), whereas the gas includes all the rest of particles. 

Decomposing also the Newton potential into the gravitational
potential of the soliton and perturbations, $\Phi=\Phi_s+\phi$,
substituting it into \cref{Saxion} and keeping only terms containing
perturbations, we obtain the gas action, 
\be
\label{Sgas}
S_g=\int dt d^3x\,\bigg(i\psi^*_g\d_t\psi_g
+\frac{\psi_g^*\Delta\psi_g}{2m}-m\Phi_s|\psi_g|^2
+\frac{\phi\Delta\phi}{8\pi G}
-m\psi_s^*\,\phi\psi_g-m\psi_s\,\phi\psi_g^*
-m\phi|\psi_g|^2\bigg).
\ee
In deriving this expression we have used that the soliton fields
$\psi_s$, $\Phi_s$ satisfy the Schr\"odinger--Poisson
equations. Following the rules of quantum field theory, we promote
$\psi_g$ and $\phi$ to second-quantized fields, whereas $\psi_s$,
$\Phi_s$ are treated as c-valued background. 
The terms linear in $\psi_g$ break the phase-rotation symmetry of the
axion gas, $\psi_g\mapsto \psi_g\e^{i\alpha}$, and therefore lead to
non-conservation of gas particles. Of course, the total number
of non-relativistic axions is conserved, meaning that the particles
from the gas go into the soliton and vice versa. The last term in
\cref{Sgas} preserves the gas particle number and describes
interactions of axions in the absence of soliton. It is responsible
for the kinetic relaxation in a homogeneous gas
\cite{Levkov:2018kau,Chavanis:2020upb}. 

\begin{figure}[tb]
\begin{center}
\begin{fmffile}{absorption1}
\parbox{70pt}{
\begin{fmfgraph*}(70,70)
\fmfpen{thick}
\fmfleft{l1,l2}
\fmfright{r1,r2}
\fmflabel{${\cal E}_1$}{l1}
\fmflabel{${\cal E}_2$}{l2}
\fmflabel{${\cal E}_3$}{r2}
\fmflabel{${\cal E}_s$}{r1}
\fmf{plain}{l1,b1}
\fmf{plain}{l2,b2}
\fmf{photon,label=$\k$}{b1,b2}
\fmf{plain}{b2,r2}  
\fmf{dashes}{b1,r1}     
\end{fmfgraph*}
}
\end{fmffile}
\quad
+
\quad
\begin{fmffile}{absorption2}
\parbox{70pt}{
\begin{fmfgraph*}(70,70)
\fmfpen{thick}
\fmfleft{l1,l2}
\fmfright{r1,r2}
\fmflabel{${\cal E}_2$}{l1}
\fmflabel{${\cal E}_1$}{l2}
\fmflabel{${\cal E}_3$}{r2}
\fmflabel{${\cal E}_s$}{r1}
\fmf{plain}{l1,b1}
\fmf{plain}{l2,b2}
\fmf{photon,label=$\k$}{b1,b2}
\fmf{plain}{b2,r2}  
\fmf{dashes}{b1,r1}     
\end{fmfgraph*}
}
\end{fmffile}
\qquad
\qquad
\begin{fmffile}{creation1}
\parbox{70pt}{
\begin{fmfgraph*}(70,70)
\fmfpen{thick}
\fmfleft{l1,l2}
\fmfright{r1,r2}
\fmflabel{${\cal E}_2$}{r2}
\fmflabel{${\cal E}_1$}{r1}
\fmflabel{${\cal E}_3$}{l2}
\fmflabel{${\cal E}_s$}{l1}
\fmf{plain}{r1,b1}
\fmf{plain}{l2,b2}
\fmf{photon,label=$\k$}{b1,b2}
\fmf{plain}{b2,r2}  
\fmf{dashes}{b1,l1}     
\end{fmfgraph*}
}
\end{fmffile}
\quad
+
\quad
\begin{fmffile}{creation2}
\parbox{70pt}{
\begin{fmfgraph*}(70,70)
\fmfpen{thick}
\fmfleft{l1,l2}
\fmfright{r1,r2}
\fmflabel{${\cal E}_1$}{r2}
\fmflabel{${\cal E}_2$}{r1}
\fmflabel{${\cal E}_3$}{l2}
\fmflabel{${\cal E}_s$}{l1}
\fmf{plain}{r1,b1}
\fmf{plain}{l2,b2}
\fmf{photon,label=$\k$}{b1,b2}
\fmf{plain}{b2,r2}  
\fmf{dashes}{b1,l1}     
\end{fmfgraph*}
}
\end{fmffile}
~~~~~\\
~~~~~\\
~~~~~\\
$(a)$\qquad\qquad\qquad\qquad\qquad
$(b)$\qquad\qquad\qquad\qquad\qquad\quad
$(c)$\qquad\qquad\qquad\qquad\qquad
$(d)$
\end{center}
\caption{Feynman diagrams describing absorption ($a,b$) and emission
  ($c,d$) of a particle by the soliton interacting with axion gas. 
Solid lines correspond to gas particles, dashed line corresponds to
the soliton, and wavy line --- to the Newtonian interaction. 
The time direction is from left to right. 
The
labels on the external legs represent the energies of the scattered
states, whereas $\k$ is the momentum exchange.    
\label{fig:scattering}}
\end{figure}
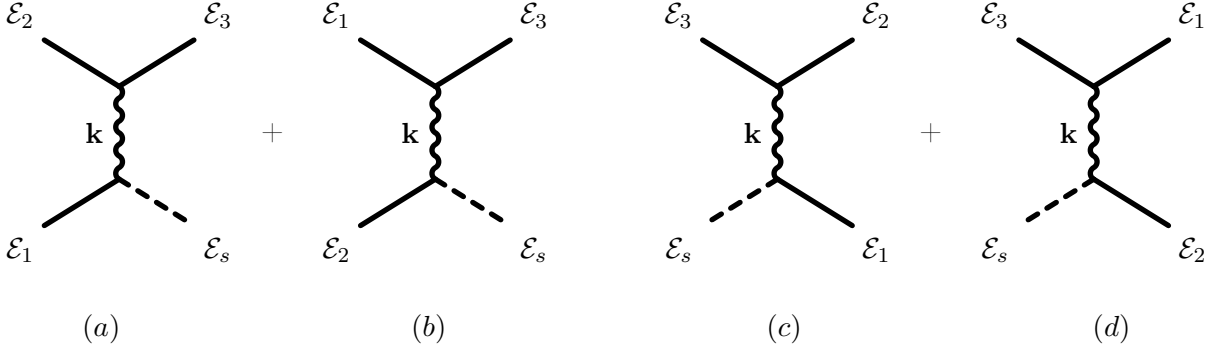

Due to energy conservation, a particle can be absorbed or emitted by
the soliton only if it exchanges energy with another particle from the
gas. This leads us to consider the process $g+g\to g+s$ when two gas
particles scatter on each other and one of them merges into the
soliton, as well as the inverse process $s+g\to g+g$ when a particle
hits the soliton and kicks out another particle. 
The Feynman diagrams for
these processes are shown in \cref{fig:scattering}. Solid straight
lines represent the gas particles, whereas dashed line corresponds
to the soliton. Wavy line stands for the ``propagator'' of the
Newton potential which is proportional to the inverse of Laplacian. In
the approximation of infinite box size it reads,
\be
\label{Newtprop}
\begin{fmffile}{Newtprop}
\parbox{85pt}{
\begin{fmfgraph*}(50,50)
\fmfpen{thick}
\fmfleft{l1}
\fmfright{r1}
\fmflabel{$(t,\x)$}{l1}
\fmflabel{$(t',\x')$}{r1}
\fmf{photon}{l1,r1}  
\end{fmfgraph*}
}
\end{fmffile}
=-i\,4\pi G\,\delta(t-t')\int \frac{[d\k]}{k^2} \e^{i\k(\x-\x')}\;, 
\ee  
where we have introduced a shorthand notation for the integration measure
\be
\label{measure}
[d\k]\equiv \frac{d^3 k}{(2\pi)^3}\; .
\ee
Combining it with the vertices implied by the action
(\ref{Sgas}), we obtain the amplitude for the diagram $(a)$ in
\cref{fig:scattering}, 
\be
\label{M1s23}
A_{1s,23}
=(2\pi)\delta(\E_1+\E_2-\E_3-\E_s)\, (4\pi Gm^2)
\int\frac{[d\k]}{k^2} V_{1s}(\k) V_{23}(-\k)\;,
\ee
with the vertex form factors
\be
\label{Vs}
V_{1s}(\k)=\int d^3x\,\psi_1(\x)\chi(|\x|)\e^{i\k\x}\;,\qquad
V_{23}(\k)=\int d^3x\,\psi_2(\x)\psi_3^*(\x)\e^{i\k\x}\;,
\ee
where 
$\psi_i(\x)$, $i = {1,2,3}$, are the wavefunctions of the states with
energies $\E_i$. The diagram $(b)$ is obtained simply by interchanging
the particles $1$ and $2$, so the total absorption amplitude is
$A_{1s,23}+A_{2s,13}$. The emission process --- diagrams $(c,d)$ in
\cref{fig:scattering} --- is described by the complex conjugate
amplitude $A_{1s,23}^*+A_{2s,13}^*$.   

The probability that two particles 1 and 2 
scatter in the way that one of them merges into soliton in unit time 
is given by
the usual formula,
\be
\frac{dp_{12\to 3s}}{dt}=(2\pi)\delta(\E_1+\E_2-\E_3-\E_s)\,
|A_{1s,23}'+A_{2s,13}'|^2\;, 
\ee
where prime denotes the amplitudes stripped off the energy
$\delta$-function,
\be
\label{Astrip}
A_{1s,23}'=(4\pi Gm^2)
\int\frac{[d\k]}{k^2} V_{1s}(\k) V_{23}(-\k)\;,
\ee
and similarly for $A_{2s,13}'$. To obtain the change in the soliton
mass, we have to subtract the rate of the inverse process and 
sum over all states in the gas weighting them with
the occupation numbers $f_i$. The weighting takes into account the
effect of the Bose enhancement due to non-zero occupation numbers of
the initial and final states. This yields,
\begin{align}
\Gamma_s&=\frac{m}{M_s}\times\frac{1}{2}
\!\sum_{\text{states 1,2,3}} 
\!\!\!\!\!(2\pi) \delta(\E_1\!+\!\E_2\!-\!\E_3\!-\!\E_s)
\big(f_1 f_2 (1\!+\!f_3)-(1\!+\!f_1)(1\!+\!f_2)f_3\big)
|A'_{1s,23}+A'_{2s,13}|^2\notag\\ 
&\simeq \frac{m}{2M_s} 
\sum_{\text{states 1,2,3}} 
\!\!(2\pi) \delta(\E_1\!+\!\E_2\!-\!\E_3\!-\!\E_s)
\big(f_1 f_2 - f_1 f_3-f_2 f_3\big)|A'_{1s,23}+A'_{2s,13}|^2\;,
\label{eq:NsGen}
\end{align}
where the factor $1/2$ has been inserted to avoid double-counting
the pairs of states related by the interchange of particles 1 and
2. In going to the second line we used that the occupation
numbers are large and kept only the leading terms quadratic in $f_i$.
Equation
(\ref{eq:NsGen}) represents the key result of this subsection. It
describes the evolution of the soliton mass for arbitrary distribution
of the gas particles. 

To proceed, we assume that the gas distribution far away from the
soliton is controlled by a single characteristic momentum $k_g$ as,
for example, in the case of the Maxwellian gas (\ref{fgas}). For the
bound states localized near the soliton, the occupation numbers can,
in principle, also depend on the soliton properties. These, as
discusses in \cref{sec:Maxwell}, are determined by a single parameter
$k_s$. Thus, we write an Ansatz,
\be
\label{fAns}
f_i=\frac{\rho_g}{mk_g^3}\; 
u\bigg(\frac{m\E_i}{k_g^2},\frac{k_g}{k_s}\bigg)\;,
\ee   
where $\rho_g$ is the density of the gas far away from the soliton,
and $u$ is a dimensionless function.
Next, it is convenient to rescale the coordinates, momenta, energies
and wavefunctions to units associated with the soliton,
\be
\x=\boldsymbol{\xi}/{k_s}~,~~~\k={k_s} \boldsymbol{\kappa}
~,~~~\E_i=\ve_i \frac{k_s^2}{m}~,~~~
\psi_i(\x)=k_s^{3/2}\vf_i({k_s} \x)\;.
\label{eq:rescale}
\ee 
Substituting these rescaled variables into eqs.~(\ref{Vs}),
(\ref{Astrip}), (\ref{eq:NsGen}) we obtain,
\be
\label{Gammagamma}
\Gamma_s=\frac{(4\pi G)^2m^3\rho_g^2}{k_g^6}\,\gamma_s(\nu)\;,
\ee
where $\nu=k_g/k_s$ is the parameter introduced in \cref{eq:alpharatio}. The
dimensionless function $\gamma_s(\nu)$ is computed by summing over the states in
the background of the standard soliton of \cref{sec:Maxwell},
\be
\label{gammadef}
\gamma_s(\nu)=\frac{\pi}{\mu_0}
\sum_{\text{states 1,2,3}} 
\!\!\delta(\ve_1\!+\!\ve_2\!-\!\ve_3\!-\!\ve_0)\;
\big(u_1 u_2 - u_1 u_3-u_2 u_3\big)|{\cal A}'_{1s,23}+{\cal A}'_{2s,13}|^2\;,
\ee
where $\ve_0$, $\mu_0$ are numerical coefficients 
quoted in \cref{standnums} and 
$u_i\equiv u(\ve_i/\nu^2,\nu)$ are rescaled occupation
numbers. For the rescaled amplitudes we have
\begin{gather}
\label{dimlessA}
{\cal A}'_{1s,23}=\int\frac{[d\boldsymbol{\kappa}]}{\kappa^2}
{\cal V}_{1s}(\boldsymbol{\kappa})
{\cal V}_{23}(-\boldsymbol{\kappa})~,\\
{\cal V}_{1s}(\boldsymbol{\kappa})=
\int d^3\xi\,\vf_1(\boldsymbol{\xi}) 
\chi_0(\xi)\e^{i\boldsymbol{\kappa\xi}}~,~~~~~~
{\cal V}_{23}(\boldsymbol{\kappa})=
\int d^3\xi\,\vf_2(\boldsymbol{\xi}) 
\vf_3^*(\boldsymbol{\xi})\e^{i\boldsymbol{\kappa\xi}}\;,
\label{dimlessVs}
\end{gather}
where $\chi_0(\xi)$ is the standard soliton profile. 
In \cref{sec:simulation} we extract the function $\gamma_s(\nu)$ from numerical
simulations, whereas in the rest of this section 
we estimate it analytically for the cases of
light and heavy solitons in Maxwellian gas.  

Before moving on, let us comment on the structure of the eigenfunctions
in the soliton background which enter into the calculation of the
soliton growth rate through the form factors (\ref{Vs}) or
(\ref{dimlessVs})
 (the details
will be presented in a forthcoming publication
\cite{soliton2}). First, it is clear from the third term in the action
(\ref{Sgas}) that the wavefunctions will be affected by the
soliton gravitational potential $\Phi_s$. While this effect is small for
highly excited unbound states with energies $\E_i\gg |\E_s|$,
it becomes important for the states with
$\E_i\lesssim |\E_s|$ and gives rise to a discrete spectrum of bound
states. Second, an additional 
modification of the eigenfunctions comes from the term
$-m\psi_s^*\,\phi\psi_g$ and its complex
conjugate in \cref{Sgas}. These terms bring qualitatively new features by mixing
positive and negative frequencies in the eigenvalue equation
\cite{Guzman:2004wj,soliton2}. As a result, the eigenmodes contain
both positive and negative frequency components
which can be interpreted as consequence of the 
Bogoliubov transformation required to diagonalize the Hamiltonian in
the presence of the condensate \cite{pitaevskii2016bose}. 
The negative-frequency part is
significant for low lying modes and cannot be discarded. In
particular, it is
crucial for the existence of zero-energy excitations
required by the spontaneously broken 
translation symmetry. 
On the other hand, for the modes of the continuous spectrum 
the negative-frequency component is
essentially negligible. 

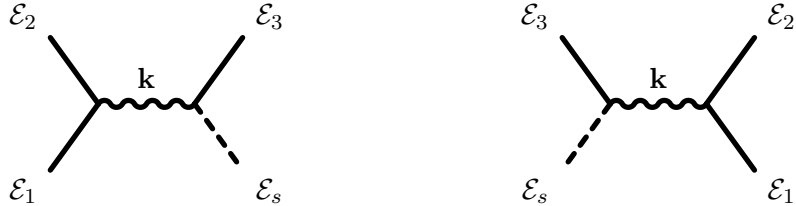
\begin{figure}[htb]
\begin{center}
~\vspace{0.5cm}\\
\begin{fmffile}{s-channel-abs}
\parbox{90pt}{
\begin{fmfgraph*}(90,50)
\fmfpen{thick}
\fmfleft{l1,l2}
\fmfright{r1,r2}
\fmflabel{${\cal E}_1$}{l1}
\fmflabel{${\cal E}_2$}{l2}
\fmflabel{${\cal E}_3$}{r2}
\fmflabel{${\cal E}_s$}{r1}
\fmf{plain}{l1,b1}
\fmf{plain}{l2,b1}
\fmf{photon,label=$\k$,label.side=left}{b1,b2}
\fmf{plain}{b2,r2}  
\fmf{dashes}{b2,r1}     
\end{fmfgraph*}
}
\end{fmffile}
\qquad\qquad\qquad\qquad
\begin{fmffile}{s-channel-emit}
\parbox{90pt}{
\begin{fmfgraph*}(90,50)
\fmfpen{thick}
\fmfleft{l1,l2}
\fmfright{r1,r2}
\fmflabel{${\cal E}_1$}{r1}
\fmflabel{${\cal E}_2$}{r2}
\fmflabel{${\cal E}_3$}{l2}
\fmflabel{${\cal E}_s$}{l1}
\fmf{plain}{r1,b2}
\fmf{plain}{r2,b2}
\fmf{photon,label=$\k$}{b1,b2}
\fmf{plain}{b1,l2}  
\fmf{dashes}{b1,l1}     
\end{fmfgraph*}
}
\end{fmffile}
\end{center}
\caption{$s$-channel diagrams for absorption (left) and emission
  (right) of a particle by the soliton arising due to mixing between
  positive and negative frequency modes. This contribution is
  subdominant compared to the diagrams in \cref{fig:scattering}
for the kinematic configurations studied in this paper.   
\label{fig:s-channel}}
\end{figure}

The admixture of negative frequencies admits, in addition to the
diagrams considered above, an $s$-channel exchange
shown in \cref{fig:s-channel}. In principle, the corresponding
amplitude $A_{3s,12}'$ should be included in the calculation of the
scattering rate. This amplitude is, however, subdominant for most
kinematic configurations. It is proportional to the negative
frequency component of particle 1 or 2 and thus 
is negligible when these
particles are unbound. In other cases, 
like in the case of light soliton studied below, it
is suppressed by the hard momentum transfer in the propagator. We do
not consider this diagram in what follows.

\subsection{Light soliton}
\label{sec:exactwf}

Calculation of $\gamma_s(\nu)$ is challenging in general. The
task simplifies for the case $\nu\gg 1$ which corresponds to
{\it light soliton} as defined in \cref{sec:Maxwell}. The typical
momentum of particles in the gas in this case is much larger than the
momentum of particles in the soliton. In other words, the soliton is
colder than the gas.

Let us understand which kinematical region gives the dominant
contribution into the sum in \cref{gammadef}. To this aim, consider
the amplitude (\ref{dimlessA}) and take the particles 2 and 3 to be
typical particles in the gas. Since their energies are much higher that
the soliton binding energy, their wavefunctions are well described by
plane waves with momenta $\boldsymbol{\kappa}_2$,
$\boldsymbol{\kappa}_3$ which are of order $\nu$. Substituting these
into the vertex ${\cal V}_{23}$ we obtain,
\be
\label{V23light}
{\cal
  V}_{23}(\boldsymbol{-\kappa})=(2\pi)^3\delta(\boldsymbol{\kappa}_2
-\boldsymbol{\kappa}_3-\boldsymbol{\kappa})\;,
\ee    
and hence the amplitude 
\be
\label{M1s23-1}
{\cal A}_{1s,23}'=\frac{{\cal
    V}_{1s}(\boldsymbol{\kappa})}{\kappa^2}~,~~~~~
\boldsymbol{\kappa}=\boldsymbol{\kappa}_2-\boldsymbol{\kappa}_3 \;.
\ee
The denominator enhances the amplitude for soft momentum
exchange. However, the exchange cannot be arbitrarily small since the
matrix element ${\cal V}_{1s}(\boldsymbol{\kappa})$ vanishes at
$\boldsymbol{\kappa}=0$ due to orthogonality of the wavefunctions
$\vf_1$ and $\chi_0$. It can be further shown \cite{soliton2} that a
linear in $\kappa$ contribution also vanishes as a consequence of
(spontaneously broken) translation invariance. Thus,
\be  
\label{V1sk2}
{\cal
  V}_{1s}(\boldsymbol{\kappa})\sim \kappa^2
\ee
and the pole in the amplitude cancels
out. We conclude that the amplitide is  
maximal at $\kappa\sim 1$
where it is of order $1$. The corresponding wavefunction $\vf_1$ must
be one of the low-lying states with characteristic energy and momentum
$|\ve_1|, \kappa_1\sim 1$. Notice that the amplitude obtained by the
interchange of particles 1 and 2 for the same kinematics is
suppressed,
\be
\label{M2s13-2}
{\cal A}_{2s,13}'=\frac{{\cal
    V}_{2s}(\boldsymbol{\kappa}_1-\boldsymbol{\kappa}_3)}{|\boldsymbol{\kappa}_1-\boldsymbol{\kappa}_3|^2}\sim \frac{1}{\kappa_3^2}\sim\frac{1}{\nu^2}\;.
\ee 

We now return to the expression (\ref{gammadef}) and rewrite it in the
following form,
\be
\begin{split}
\gamma_s(\nu)=
\frac{\pi}{\mu_0}\sum_{\text{states 1,2,3}} 
\delta(\ve_1+\ve_2-\ve_3&-\ve_0)\big[2u_1(u_2-u_3)|{\cal A}'_{1s,23}|^2
-2u_2u_3
|{\cal A}'_{1s,23}|^2\\
&+(u_1u_2-u_1u_3-u_2u_3)({\cal A}'_{1s,23}{\cal A}'^*_{2s,13}+\text{h.c})\big]\;.
\end{split}
\label{eq:dN_F2}
\ee
For the preferred kinematics,
the first term in brackets is small. Indeed, using the Maxwell
distribution for the unbounded states we obtain,
\be
u_2-u_3 = u_2 \, 
\big(1-\e^{-2(\ve_3-\ve_2)/\nu^2}\big)=
u_2  \, \big(1-\e^{- 2 (\ve_1-\ve_0)/\nu^2}\big)
\approx u_2\frac{2(\ve_1-\ve_0)}{\nu^2} = O(\nu^{-2})\;,
\ee 
where in the second equality we used the energy conservation.
The terms in the second line in \cref{eq:dN_F2} are also suppressed
due to \cref{M2s13-2}. Thus, up to corrections of order
$O(\nu^{-2})$, we have
\be
\label{gammalight1}
\gamma_s(\nu)=-\frac{2\pi}{\m_0}\sum_{\text{state 1}}
\int[d\boldsymbol{\kappa}_2][d\boldsymbol{\kappa}_3]
\delta\Big(\ve_1-\ve_0+\tfrac{\kappa_2^2}{2}-\tfrac{\kappa_3^2}{2}\Big)
(4\pi)^3\e^{-(\kappa_2^2+\kappa_3^2)/\nu^2}
\frac{|{\cal V}_{1s}(\boldsymbol{\kappa}_2-\boldsymbol{\kappa}_3)|^2}{|\boldsymbol{\kappa}_2-\boldsymbol{\kappa}_3|^4}\;.
\ee
Two comments are in order. First, we observe that $\gamma_s(\nu)$ is
negative. Recalling that it multiplies the rate of the soliton mass
change, \cref{Gammagamma}, we conclude that the mass of a light
soliton decreases --- it {\em evaporates}. Second, the expression
(\ref{gammalight1}) does not depend on the occupation number of the
low-lying state 1.  
This is a nice property.
Particles from the low-lying energy levels are further
upscattered by the gas and eventually become 
unbound. 
Calculation of the occupation numbers of these levels presents a
nontrivial task.  
Fortunately, we don't need to know them to determine the soliton
evaporation rate in the leading order. 

The next steps include changing the integration variables to
$\boldsymbol{\kappa}=\boldsymbol{\kappa}_2-\boldsymbol{\kappa}_3$ and 
$\boldsymbol{\kappa}_+=(\boldsymbol{\kappa}_2+\boldsymbol{\kappa}_3)/2$ and
performing the integration over $\boldsymbol{\kappa}_+$. Discarding
suppressed terms, we obtain that $\gamma_s$ is proportional to $\nu^2$
with a numerical coefficient equal to a certain weighted 
sum over states in the standard
soliton background,
\be
\label{F2new}
\gamma_s(\nu)=-C_{ls}\,\nu^2~,~~~~~~C_{ls}=\frac{8\pi^2}{\mu_0}
\sum_{\text{state 1}}\int \frac{[d\boldsymbol{\kappa}]}{\kappa^5}|{\cal
  V}_{1s}(\boldsymbol{\kappa})|^2\;. 
\ee 
Despite an 
apparent pole of the integrand
at $\kappa\to 0$, the coefficient $C_{ls}$ is finite due to the property
(\ref{V1sk2}). 
Numerical evaluation gives
\cite{soliton2},
\be
\label{Clsnum}
C_{ls}= 4.3\pm 0.2\;.
\ee

To summarize, the light solitons evaporate. 
The change of the soliton mass 
is dominated by the process of $g+s \to 
g+ g$, with gas particles kicking off axions from the soliton. 
By considering the soft momentum exchange, we have obtained
the leading term in the function $\gamma_s(\nu)$ in the 
evaporation rate, which is proportional to $\nu^2$ with an order-one
coefficient. 

It is instructive to compare the time scale of evaporation
$|\Gamma_s|^{-1}$ with the relaxation time in the gas (\ref{trelax}).
We see that evaporation is faster than relaxation if $\nu$ exceeds the
critical values
\be
\label{nucrit}
\nu_c=\sqrt{\frac{3\pi\,\ln(k_gL)}{4\sqrt 2\, b\, C_{ls}}}\simeq 1.5\;, 
\ee
where we have used $\ln(k_gL)\sim 5$.
This is close to the threshold for soliton evaporation
found in numerical simulations, see \cref{sec:simulation}. For 
$\nu>\nu_c$ the relaxation in the gas can be neglected and our
assumption of the stability of the Maxwell distribution is well
justified.

\subsection{Heavy soliton}
\label{sec:heavysoliton}

In this section we consider the opposite limit $\nu\ll 1$ corresponding
to {\it heavy} or {\it hot} soliton. The analysis in this case is more
complicated, so we content ourselves with semi-qualitative discussion
focusing on the overall scaling of the growth rate function $\gamma_s$ 
with $\nu$. A more detailed study is left for future.

For heavy soliton, the typical energy of gas particles is much smaller
than the soliton binding energy which in our dimensionless 
units is of order one. Then the process with kicking-off
particles from the soliton shown on the right of \cref{fig:scattering}
is strongly suppressed since it requires from particle $3$ to have
order-one energy. We are left with the absorption process given by the
diagrams $(a,b)$ on \cref{fig:scattering} and corresponding to the
term proportional to $u_1u_2$ in \cref{gammadef}. This already allows
us to conclude that 
the heavy soliton grows at a strictly positive
rate, thereby excluding the possibility of a kinetic
equilibrium between the soliton and the gas. 

Particles 1 and 2 that participate in the absorption process can belong
either to unbound or to bound states. A problem
arises because the occupation numbers of the bound states are
unknown. In a complete treatment, they must be determined
self-consistently from the solution of the Boltzmann equation in the
gas. Such analysis is beyond the scope of this paper. Below we 
focus on the contribution into $\gamma_s(\nu)$ coming from the
processes when both states 1 and 2 are unbound, assuming that it
correctly captures the scaling of the full result with $\nu$. We
stress that this assumption must be verified by a detailed study which
we postpone to future. We further assume that the
occupation numbers of the unbound states are Maxwellian.    

Even for unbound sates, the wavefunctions are significantly modified
by the long-range Newtonian potential of the soliton which in 
the dimensionless units has the form,
\begin{equation}
   U(\xi) =  - \frac{\mu_0 }{4\pi \xi} \equiv 
  - \frac{\beta}{\xi} 
 \ .
 \label{eq:Ur}
\end{equation}
We can capture its effect by approximating the exact eigenfunctions
with the Coulomb
wavefunctions,
\begin{equation} 
\label{Coulombwave}
   \vf_{\boldsymbol{\kappa}} (\boldsymbol{\xi} ) =
\e^{i(\beta/\kappa)(\ln{\beta/\kappa}-1)+i\pi/4}\,
 \varGamma \bigg( 1 - i\frac{\beta}{\kappa}\bigg ) 
\, \e^{ \pi\beta /(2\kappa)}\, \e^{i \boldsymbol{ \kappa\xi} } \, 
           {}_{1}F_{1} \bigg( i \frac{\beta}{\kappa} ; 1; i (\kappa \xi - 
i \boldsymbol{\kappa\xi} ) \bigg)  \ ,
\end{equation} 
where $\varGamma$ stands for the gamma-function and 
${}_{1}F_{1}$ is the confluent hypergeometric (Kummer)
function. This solution describes a scattered wave with initial momentum
$\boldsymbol{\kappa}$. Note that, compared to the standard definition, 
we have added a phase in
\cref{Coulombwave} for later
convenience.  

For modes with small asymptotic momenta the eigenfunctions simplify, 
\begin{align}
\label{Coulombsoft}
   \vf_{\boldsymbol{\kappa}} (\boldsymbol{\xi} )  
\to   \sqrt{ \frac{2 \pi\beta}{\kappa} } \, 
         J_0 \big(2 \sqrt{ \beta ( \xi -  {\bf n} \boldsymbol{\xi} ) } \big)   
      \equiv \frac{1}{\sqrt{\kappa}} \, \hat{ \varphi}_{\bf n}
      (\boldsymbol{\xi}) ~,~~~~~\kappa\ll 1    \ ,
\end{align}
where ${\bf n}=\boldsymbol{\kappa}/\kappa$ is the unit vector in the
direction of momentum. We observe that the dependence on the absolute
value of momentum factorizes. Note that the eigenfunctions get
enhanced at $\kappa\to 0$ which reflects the focusing effect of the
Coulomb field. Note also that, despite the small momentum at infinity,
the eigenfunctions oscillate with order-one period at $\xi\sim 1$,
consistent with the fact that particles accelerate to an order-one momentum
in the vicinity of the soliton.

We now use \cref{Coulombsoft} for the gas particles 1 and 2  (but not
for the particle 3 which has $\kappa_3\sim 1$). This yields for the
amplitude, 
\bseq
\label{eq:heavy_V23}
\begin{align}
  & {\cal V}_{1s} (\boldsymbol{\kappa} ) =
 \frac{1}{\sqrt{\kappa_1}} \int d^3\xi\,\hat\vf_{{\bf
   n}_1}(\boldsymbol{\xi})
\chi_0(\xi)\e^{i\boldsymbol{\kappa\xi}}
            \equiv\frac{1} { \sqrt{\kappa_1}} 
\hat{\cal V}_{1s}(\boldsymbol{\kappa} )\;,\\
 &  {\cal V}_{23} (\boldsymbol{\kappa} ) =
 \frac{1}{\sqrt{\kappa_2}} \int d^3\xi\,\hat\vf_{{\bf
   n}_2}(\boldsymbol{\xi})
\vf_{\boldsymbol{\kappa}_3}^*(\boldsymbol{\xi})\e^{i\boldsymbol{\kappa\xi}}
            \equiv\frac{1} { \sqrt{\kappa_2}} 
\hat{\cal V}_{23}(\boldsymbol{\kappa} )\;,\\
&{\cal A}'_{1s,23}=\frac{1}{\sqrt{\kappa_1\kappa_2}}
\int\frac{[d\boldsymbol{\kappa}]}{\kappa^2} 
\hat{\cal V}_{1s}(\boldsymbol{\kappa} )
\hat{\cal V}_{23}(-\boldsymbol{\kappa} )
\equiv
\frac{1}{\sqrt{\kappa_1\kappa_2}} {\hat{\cal A}}'_{1s,23}\;,
\end{align}
\eseq
where the hatted quantities do not depend on the absolute values of
the momenta $\kappa_1$, $\kappa_2$. We substitute this into
the expression for $\gamma_s$ and, upon neglecting $\ve_1$, $\ve_2$ in
the energy $\delta$-function, perform the integration over $\kappa_1$,
$\kappa_2$. In this way we obtain,
\be
\label{gammasu}
\gamma_s^{(u)}(\nu)=\frac{\nu^4}{(2\pi)^2\mu_0}
\int d{\bf n}_1 d{\bf n}_2[d\boldsymbol{\kappa}_3]\;
\delta\bigg(\frac{\kappa_3^2}{2}+\ve_0\bigg)
|\hat{\cal A}'_{1s,23}+\hat{\cal A}'_{2s,13}|^2\;,
\ee
where the superscript $(u)$ is to remind that we consider only the
contribution from unbound states. All quantities inside the integral
are $\nu$-independent. Thus we conclude that $\gamma_s^{(u)}$ scales
as the fourth power of $\nu$. Assuming that this also holds for the
full contribution we write,
\be
\label{eq:heavyS_rate}
\gamma_s(\nu)=C_{hs} \nu^4~,~~~~C_{hs}>0,~~~~~\text{at}~\nu\to 0\;.
\ee
This implies that the soliton growth slows down with the
increase of the soliton mass.

We do not attempt to estimate the numerical coefficient $C_{hs}$. 
As already mentioned, this would require inclusion of the
bound state contribution which is beyond our present scope. Another
caveat comes from the fact that the time scale of the heavy soliton
growth $\Gamma_s^{-1}$ happens to be parametrically longer than the
gas relaxation time (\ref{trelax}). On these time scales the gas
distribution may evolve away from Maxwellian which we assumed in our
derivation.\footnote{As discussed below, numerical simulations suggest
that Maxwell distribution may still be a good approximation, but this
question requires further study.} 
Thus, the formula (\ref{eq:heavyS_rate}) should be taken
with a grain of salt. Its comparison with the results of simulations
is discussed in the next section.

\section{Wave Simulations}
\label{sec:simulation}

In this section we present our numerical simulations. We first
describe the setup. Then we provide three typical examples of
simulation runs for heavy, intermediate and light solitons and
introduce the procedure which we use to measure the soliton growth
rate. 
Finally,
we assemble 195 individual simulation runs to extract the soliton
growth/evaporation rates and compare them to the theoretical
predictions of the previous section. 
We focus here on the main suit of simulations where in each run we
assign a single soliton surrounded by Maxwellian axion gas as the
initial conditions. In \cref{sec:levkov} we also report the
simulations without the initial
soliton where it forms dynamically from the axion gas, as in
Ref.~\cite{Levkov:2018kau}.

\subsection{Setup}
\label{sec:setup}

\subsubsection*{Evolution}
We use the scaling transformation (\ref{eq:scaling}) to convert the
Schr\"odinger--Poisson equations into the following dimensionless form,
\bseq
\label{eq:dimensionlessEq}
\begin{align}
\label{eqSchrSim}
&i\d_t\tilde\psi +\frac{1}{2} \Delta\tilde\psi - \tilde\Phi \,
\tilde\psi=0\;, \\ 
\label{eqPoisSim}
&\Delta\tilde \Phi=|\tilde\psi|^2\;,
\end{align}
\eseq
which is equivalent to the choice of units $m=4\pi G=1$. This system
is solved on a cubic lattice of size ${\cal N}$ with periodic boundary
conditions on $\tilde \psi$ and $\tilde\Phi$. We use the residual
scaling symmetry to fix the lattice spacing to one, ${\rm d}x=1$. The
size of the lattice thus sets the length of the box side
and remains a free parameter. We run simulations for three different
values ${\cal N}=128,~256,~512$. In what follows we omit tildes over
dimensionless quantities.

The wavefunction is advanced by the leapfrog integration algorithm
(drift-kick-drift)~\cite{birdsall2018plasma,ChanEtal18}, 
\begin{equation}
\label{DKD}
   \psi(t+ {\rm d} t, {\bf x} )= \e^{i \, \Delta \, {\rm d} t/4} 
\cdot \e^{-i \, \Phi(t+{\rm d} t/2, {\bf x}  )  \,  
      {\rm d} t } 
            \cdot \e^{i \, \Delta \, {\rm d} t/4}  \, \psi(t, {\bf x}  )  \, . 
\end{equation}
We transform $\psi$ to the momentum space to evolve with $ \e^{i \, \Delta
  \, {\rm d} t/4}$ and $\Delta$ is converted to $-{\bf k}^2$,
while the evolution with the gravitational potential, $\e^{-i \,
  \Phi  \,  {\rm d} t }$, is performed in the real
space. Fourier components of the gravitational potential with $\k\neq
0$ are found from \cref{eqPoisSim},
\be
\Phi_\k=-\frac{(|\psi|^2)_\k}{\k^2}\;,
\ee
whereas the zero mode is set to vanish,\footnote{This 
  corresponds to an arbitrary choice of the zero-point energy in the
  Schr\"odinger equation~(\ref{eqSchrSim}).} $\Phi_{\bf k=0}=0$.
We use uniform time step 
${\rm d} t=2/\pi$ which 
is determined by the requirement that the phase difference of a
high-momentum 
mode with $k=\pi$ between consecutive time slices
does not exceed $\pi$. 
To assess the accuracy of the simulations, we monitor the total energy
of the axion field in the box,
\be
\label{Etot}
E=\frac{1}{2}\sum_\k\k^2|\psi_\k|^2+\frac{1}{2}\sum_\x \Phi(\x)|\psi(\x)|^2\;.
\ee
We have observed that the energy conservation quickly deteriorates for
heavy solitons with sizes comparable to the lattice spacing,
$r_s\sim 1$ (see \cref{sec:resolution} for details). In our 
analysis we only use runs where the energy is conserved with the
precision $\lesssim 0.1\%$.

\subsubsection*{Initial conditions for axion gas}
The gas wavefunction is set up in the initial conditions through its
Fourier decomposition,
\begin{equation}
    \psi_g(t=0,\x)=\frac{1}{\Nres^{3/2}}
\sum_\mathbf{k}a_\k \cdot \e^{i\mathbf{k}\cdot\mathbf{x}} \, ,
\end{equation}
where the absolute values of the amplitudes $a_\k$ are taken to follow
the Maxwell distribution (\ref{fgas}). To ensure that the gas modes
are well resolved on the lattice, we restrict to $k_g\leq 1$. The
phases of $a_\k$ are assigned to random numbers uniformly distributed
in the range $(0, 2\pi)$. We have repeated simulations for several
random initial phase realizations and have found that the choice of
realization does not affect our results. The mean gas density $\rho_g$ and its
total mass $M_g$ can be deduced as
\begin{equation}
    \rho_g= \frac{1}{\Nres^3} \, \int {\rm d}^3 \mathbf{x}\,
    |\psi(\mathbf{x}) |^2  =    \frac{ f_g k_g^3}{(4\pi)^{3/2}} \, ,  
   \qquad\qquad M_g = \rho_g \,  \Nres^3 = \frac{ f_g k_g^3 \Nres^3  } {
     (4\pi)^{3/2} } \, . 
\label{eq:rho0}
\end{equation}
The gas density is limited from above by the condition to avoid the
Jeans instability that triggers a halo formation and thereby
complicates the interpretation of simulation results. Thus, we require
the size of the simulation box to be smaller than the Jeans length
(\ref{lJeans}), which yields the condition:
\begin{equation}   
   \Nres < l_J ~~~~~\Longleftrightarrow~~~~~
   f_g \, k_g < 0.054 \, \left( \frac{\Nres }{ 128}  \right)^{-2} \ .
   \label{eqn:jeans_instability}
\end{equation}   
This puts a stringent restriction on the parameter space of the
simulations.

\subsubsection*{Initial conditions for soliton}
We superimpose
the soliton wavefunction on top of the gas wavefunction at the
beginning of the simulations.\footnote{Dynamical soliton formation from
  the gas is discussed in \cref{sec:levkov}.} 
The input soliton density profile uses the analytic fit 
(\ref{eq:rhos}) characterized by a single parameter, the half-peak radius
$r_{s}^{\rm init}$. The peak density of the fit is taken to be
\cite{Schive:2014hza},
\begin{equation}
\rho_{s,\, {\rm peak}}^{\rm init}   = \frac{2.794}{(r_{s}^{\rm init})^{4}},
\end{equation}
which is slightly lower (by less than $2\%$) than the exact value
implied by the formulas of \cref{sec:Maxwell}. This discrepancy is
negligible given other uncertainties of the simulations.  
The initial phase of the soliton wave function is set to be zero. This
choice does not change our average result 
since the phases of the axion gas are random.
We notice that the initial soliton gets 
slightly deformed after superposing on the wavefunction of axion gas, but 
this deformation has little effect on the late time evolution. 

We take $\rinit\geq 1.5$ for the soliton to be resolved on the lattice.
Periodic boundary conditions give rise to image solitons at distance
$\Nres$ from the central one. We have observed that these images can
distort the central soliton wavefunction. To avoid this distortion, we
require the soliton size to be much smaller than the box, $r_s^{\rm
  init}<0.1\,\Nres$.

\subsubsection*{Measurement of the soliton mass}
During the simulations the radius of the soliton evolves together with
its mass. We estimate $r_s$, $M_s$ at a given time using their
relation to the soliton peak density provided by the fit to the
soliton density profile,\footnote{The expression for the soliton mass
(\ref{eq:Mg}) is by $3\%$ lower for a given peak density than the value obtained
from the exact wavefunction, see \cref{sec:Maxwell}. This
error is insignificant for our analysis. Note that its effect is
opposite to the bias introduced by the interference with the axion gas
discussed below.}
\begin{equation}
r_s=1.293\, \rho_{s,\,{\rm peak}}^{-1/4}~,~~~~~~~
M_s = 25.04\, \rho_{s,\,{\rm peak}}^{1/4}.
\label{eq:Mg}
\end{equation}
Since the soliton moves through the box during simulations, the
position of its peak is unknown. We choose the maximal density in the
whole box as a proxy for the soliton peak density assuming that the
soliton is prominent within the axion gas. Note that due to interference
between the soliton and the gas, the peak density of the axion field
does not, in general, coincide with the soliton peak. Choosing the
maximal density in the box can bias our estimate of the soliton peak
density, and hence of its mass, upwards. Detailed investigation of
this bias is performed in \cref{sec:peak}. It shows that 
the bias is at most $20\%$ when the maximal density is higher than the
mean gas density by a factor of $30$ and quickly decreases for higher
density contrasts. To obtain the soliton growth rate we analyze only
the parts of the 
simulations with
$\rho_{s,\,{\rm peak}}>30\,\rho_g$. 

On the other hand, we require the mass of the soliton to be
significantly smaller
than the total mass of the gas in order to avoid 
any effects on the
soliton evolution that can arise due to a shortage of particles in the gas. We
implement this by the condition $M_s<0.5\, M_g$.

\subsubsection*{Parameter space}
%\label{subsubsec:parameter}

%
\begin{figure}[t]
\begin{center}
 \includegraphics[width=1.0\textwidth]{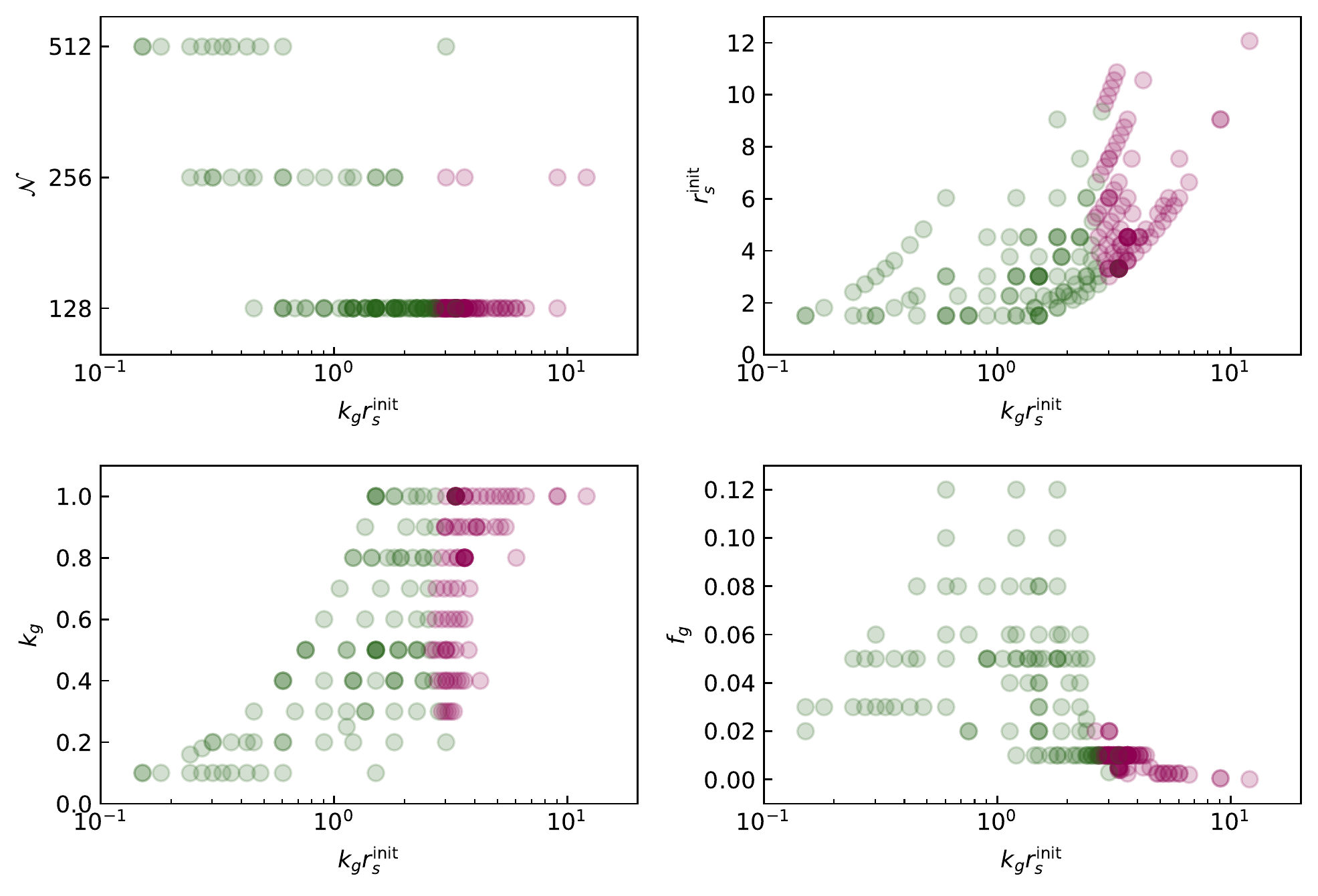}
\caption{Parameters of 195 individual simulations used in this
  work. The four-dimensional parameter space is projected on the
  directions corresponding to the box size $\Nres$, the 
soliton half-peak radius $\rinit$, and the parameters of the
Maxwell distribution of axion gas  $k_g$, $f_g$.
The horizontal axis is common to all panels and shows the product
$k_g\rinit$. 
Green circles correspond to simulations leading to soliton growth,
while red circles show the cases of soliton evaporation. 
Darker circles indicate multiple realizations of axion gas by
changing the phases in the wavefunction. 
  \label{fig:input}
}
  \end{center}
\end{figure}

Our simulations have 
four input parameters: $\Nres$, $k_g$, $f_g$, and $\rinit$, which describe
the box size, the  
momentum distribution of axion gas, and the size of soliton.
In this work, we use three box sizes, $\Nres=128$, $256$, and $512$.
For the regime of light soliton, most of the simulations are conducted
with $\Nres=128$, while for heavy solitons  
we use large boxes $\Nres=512$ in order to reach low $(k_gr_s)\sim
0.1$.
The remaining three parameters are sampled in the ranges
\be
k_g \in ( 0.1\,,\, 1)~,~~~~~f_g \in (10^{-4}\,,\, 0.12)~,~~~~~
\rinit \in (1.5\,,\,  12 )\;.
\ee
Their choice is dictated by the goal to efficiently capture the
soliton growth/evaporation within realistic simulation time, while
resolving the axion gas and the soliton on the lattice. In addition,
they are subject to constraints discussed above which we summarize
here for clarity:
\begin{itemize}
   \item[a)]  $f_g \, k_g < 0.054 \, \left(\Nres/128 \right)^{-2}$: the
     axion gas does not form a halo due to Jeans instability; 
\item[b)] $\rinit < 0.1\, \Nres$: the effect of periodic images on the
  soliton is suppressed;
  \item[c)] $\rhoc > 30\rho_g$: soliton is prominent enough to 
suppress bias in its mass measurement;
  \item[d)] $M_s < 0.5\, M_g$: soliton does not overwhelm axion waves.
\end{itemize}
Note that the conditions (a,b) are imposed on the initial
configuration, whereas the conditions (c,d) are monitored throughout
the whole duration of the simulations.
In total we have run 195 simulations with independent realizations of
random gas phases. Their parameters are shown in 
\cref{fig:input} against the product $k_g r_s^{\rm init}$ which
controls 
the physics of the soliton-gas interaction.

\subsection{Growing and evaporating solitons }
\label{sec:case}

In this section we present a case study of several simulations that
illustrate possible evolution of the soliton-gas system. We use
these examples to introduce our procedure for extraction of the soliton
growth rate. We also provide evidence that the gas distribution
remains close to Maxwellian during the simulations.

\begin{figure}[t!]
\begin{center}
\includegraphics[width=0.9\textwidth]{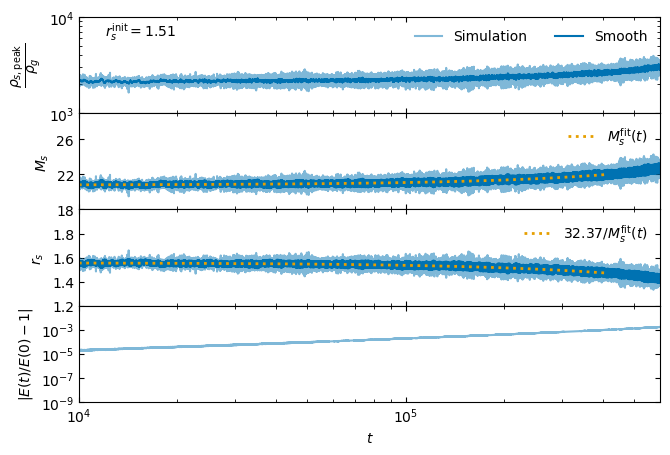}
\caption{Evolution of the soliton peak density,
 mass and radius for the case of heavy soliton ($\rinit=1.51$).  
The mass and radius are estimated from the
 peak density. Thin blue curves show the instantaneous values, 
whereas the thick curves are obtained by smoothing
with a top-hat filter. Yellow dots show the result of fitting the
soliton mass
with a quadratic polynomial.  
We also show the time dependence of the
 total energy in the simulation box used to control the precision of
 numerical calculations. The gas parameters 
 are ($\Nres=128$, $k_g = 1$, $f_g =0.01$). 
  \label{fig:cases}
}
  \end{center}
\end{figure}

\begin{figure}[ht]
\begin{center}
\includegraphics[width=0.9\textwidth]{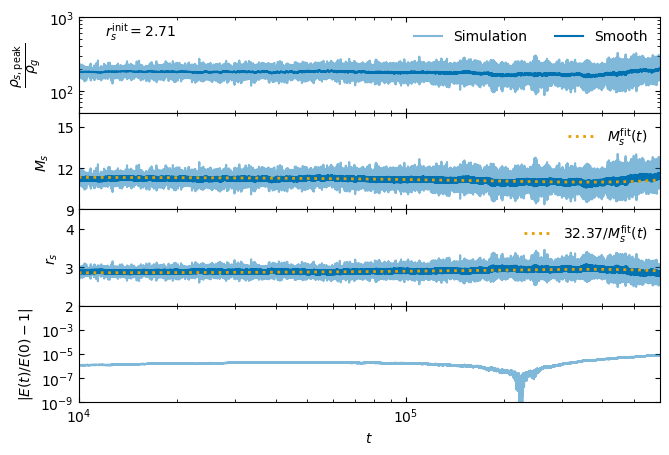}
\caption{Same as \cref{fig:cases} for the case of median soliton
($\rinit=2.71$).
  \label{fig:cases1}
}
  \end{center}
\end{figure}

\begin{figure}[ht]
\begin{center}
\includegraphics[width=0.9\textwidth]{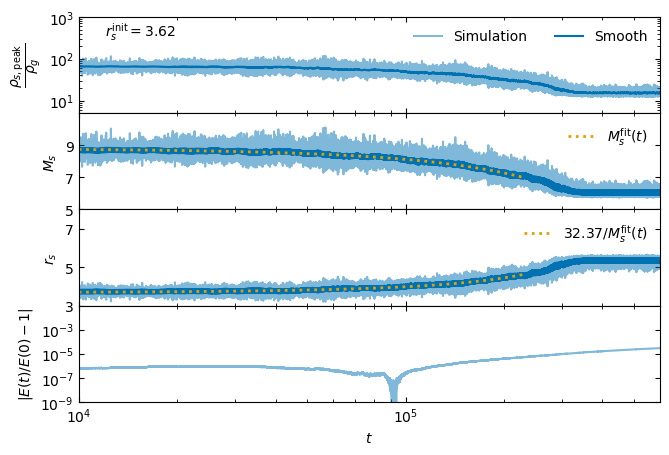}
\caption{Same as \cref{fig:cases} for the case of light soliton
($\rinit=3.62$). 
  \label{fig:cases2}
}
  \end{center}
\end{figure}

We consider three simulation runs with the same initial gas
configuration characterized by $(\Nres=128,k_g=1,f_g=0.01)$ and 
different initial soliton sizes $r_s^{\rm init}$: $1.51$ (heavy
soliton), $2.71$ (median soliton), and $3.62$ (light
soliton). Figures~\ref{fig:cases}-\ref{fig:cases2} 
show the evolution of the soliton
characteristics in the three runs. These include the soliton peak
density $\rhoc(t)$ (which we identify with the maximal density in the
box), the soliton mass $M_s(t)$ and the soliton radius $r_s(t)$. The
peak density is normalized to the mean density of the gas, whereas the
mass and radius are determined using the relations
(\ref{eq:Mg}). Clearly, the heavy soliton grows and the light soliton
evaporates which is consistent with the analysis of
\cref{sec:theory}. The 
median soliton remains approximately
unchanged indicating that the transition from growth to evaporation
occurs at $(k_gr_s)\sim 2.7$.
We also plot in figs.~\ref{fig:cases}-\ref{fig:cases2} the change in the
total energy of the axion field in the box. 
For the median and light solitons the energy is conserved with high
precision $|E(t)/E(0)-1|\lesssim 10^{-5}$ throughout the whole duration of the
simulations. For the heavy soliton, the energy exhibits a slow drift
and the error exceeds $0.1\%$ by the end of the simulations. We
associate this with the loss of spatial and temporal resolution for
heavy solitons which have small sizes $r_s\sim 1$ and high oscillation
frequencies $|\E_s|\sim 1$ (see \cref{sec:resolution} for a detailed
discussion). In our analysis we use only the portion of the simulation
where $|E(t)/E(0)-1| < 10^{-3}$.

We now describe our algorithm to extract the soliton growth rate
$\Gamma_s$. The task is complicated by strong oscillations of the
soliton peak density which are clearly visible in the plots and
translate into oscillations of the estimated soliton mass and
radius. Such oscillations have been observed in previous works
\cite{Veltmaat:2018,Eggemeier:2019jsu} 
and correspond to the normal modes of the soliton
\cite{Guzman:2004wj,Guzman19} 
with the frequency of the lowest mode $\omega\sim 0.5\,r_s^{-2}$. To
mitigate their effect, we construct running averages of the soliton
parameters by smoothing them with a top-hat function.\footnote{Note
  that we smooth $\rho_{s,\,{\rm peak}}(t)$, $M_s(t)$ and $r_s(t)$ separately.} 
We take the
width of the top-hat as a function of the initial soliton size   
$t_{\rm width}=70(\rinit)^2$ which covers about five periods of the
oscillations. The resulting smoothed dependences are shown in
figs.~\ref{fig:cases}-\ref{fig:cases2} by thick curves. 

While smoothing suppresses most of the chaotic oscillations, it still
leaves some irregularities in the time dependence of the soliton mass
that introduce significant noise when calculating its time
derivative. To further suppress this noise, we fit the smoothed mass
with an analytic function of time. We have found that a quadratic fit
is sufficient in all cases. Thus, we write  
\begin{equation}
M_s^{\rm fit}(t)=a_0+a_1t+a_2t^2 \;,   
\label{eq:Ms_fit}
\end{equation}
where $a_0$, $a_1$ and $a_2$ are fitting parameters.
The fitting time-range is determined by the following criteria:
\begin{itemize}
\item
Inside the range the soliton peak density, mass and radius satisfy the
conditions (c,d) from \cref{sec:setup};
\item
The total energy in the simulation box is conserved within precision
$|E(t) / E(0) -1| <  0.1\%$;
\item
The time duration is smaller than half of the relaxation time
(\ref{trelax}) to avoid possible changes in the gas distribution due
to kinetic relaxation \cite{Levkov:2018kau}.\footnote{In principle, 
this requirement might be too stringent 
since we observe that in the presence of a
  soliton the gas distribution remains close to Maxwellian even on
  time scales longer than the relaxation time, as will be discussed shortly.}
\end{itemize}
The best-fit values of $a_0,a_1,a_2$ for the three sample runs are
given in \cref{table:cases}. The corresponding fitted curves are shown
in figs.~\ref{fig:cases}-\ref{fig:cases2} with yellow dots. We also define the ``fitted''
soliton radius by converting it from the soliton mass 
in accordance with eqs.~(\ref{eq:Mg}),
\be
\label{rsfit}
r_s^{\rm fit}(t)\equiv\frac{32.37}{M_s^{\rm
    fit}(t)}=\frac{32.37}{a_0+a_1t+a_2t^2}\;. 
\ee
The result matches very well the smoothed dependence $r_s(t)$, see
figs.~\ref{fig:cases}-\ref{fig:cases2}. 
We have verified that an independent fit of smoothed
$r_s(t)$ with a quadratic polynomial produces essentially identical
curves, which provides a consistency check of
our procedure.

\begin{table}[t]
   \begin{center}
   \begin{tabular}{| c |c||c|c|c |    }
      \hline
          & $\rinit$ &$ a_0$ &$a_1$&$a_2$
               \\ 
   %[0.5ex]
      \hline
      \makecell{heavy soliton  }    
      & 1.51 & $20.79$&$0.283\times 10^{-5}$&$0.00239\times 10^{-10}$
      \\
      \hline
      \makecell{median soliton }   
      & 2.71 & $11.35$&$-0.203\times 10^{-5}$&$0.0282\times 10^{-10}$
       \\
      \hline
      \makecell{light soliton  }  
      & 3.62 & $8.80$&$-0.595\times 10^{-5}$&$-0.0837\times 10^{-10}$
       \\
      \hline
   \end{tabular}
   \end{center}
   \caption{Parameters of the soliton mass fit 
 for the three simulations
     shown in figs.~\ref{fig:cases}-\ref{fig:cases2}. 
The initial size of the soliton is
     $\rinit$. 
 The parameters of axion gas are
     $\Nres=128$, 
     $k_g=1$, $f_g=0.01$.}
   \label{table:cases}
\end{table}

We can now estimate the soliton growth rate substituting the fitted
time dependence of the soliton mass in the defining formula
(\ref{rate_def}), which yields,
\begin{equation}
\Gamma_s^{\rm fit}(t) =\frac{a_1+2\,a_2\,t}{a_0+a_1\,t+a_2\,t^2}\;. 
\end{equation}
We are interested in the dependence of the growth rate on the soliton
radius $r_s$. Both these quantities depend on time, so a single run 
provides a continuous set of data points
$\big(r_s^{\rm fit}(t),\Gamma_s^{\rm fit}(t)\big)$ 
sampled at different moments of
time. In view of uncertainties of our
smoothing and fitting procedure, we reduce this set to 20 data points 
$\big(r_s^{\rm fit}(t_i),\Gamma_s^{\rm fit}(t_i)\big)$, $i=1,\ldots,20$, 
evenly distributed in time within the range
of the fitting function $M_s^{\rm fit}(t)$. 
These 20 data points represent the output of a
single run. In the next subsection we combine the outputs of 195 runs
to build the cumulative dependence of the growth rate on the soliton
and gas parameters.

\begin{figure}[t]
     \centering
     \begin{subfigure}[b]{0.75\textwidth}
         \centering
         \includegraphics[width=\textwidth]{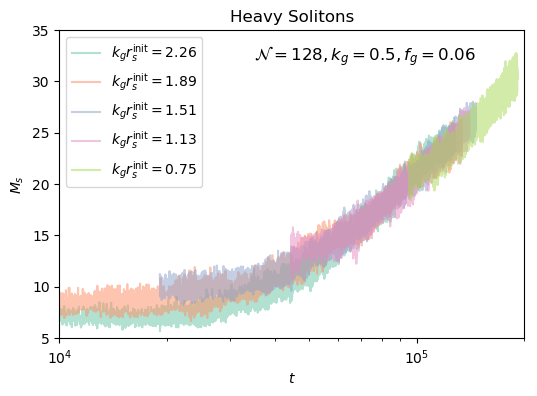}
         \label{fig:stack1}
     \end{subfigure}
     \hfill
     \begin{subfigure}[b]{0.75\textwidth}
         \centering
         \includegraphics[width=\textwidth]{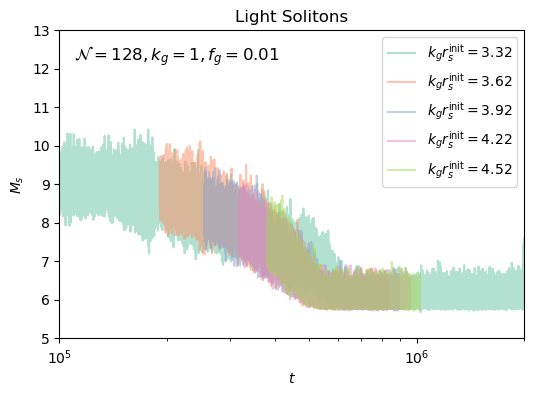}
         \label{fig:stack2}
     \end{subfigure}
\caption{
Soliton mass evolution in 
simulations with $k_g\rinit$ 
from 0.75 to 2.26 (top) and from 3.32 to 4.52 (bottom). 
By shifting the curves along the time axis we have observed that 
they can be stacked on top
of each other.
\label{fig:stack}    
} 
\end{figure}

Soliton growth rate depends on the gas distribution which can, in
principle, change during the simulations. This could lead to
incompatibility of the results read out at different moments from the
start of the runs. To verify that this is not the case, we compare the
runs that differ by the initial soliton mass, but have overlapping
soliton mass ranges spanned during the evolution. The top panel of 
\cref{fig:stack} shows the evolution of the soliton mass in five
simulations of heavy solitons with $k_g\rinit$ varying from $0.75$ to
$2.26$. The gas parameters are chosen the same in all five runs
$(\Nres=128,\,k_g=0.5,\,f_g=0.06)$. The curves have been shifted in
time until they overlap. 
We observe that the curves are well
aligned with each other. In the lower panel of \cref{fig:stack}  we
repeat the same exercise for five light soliton simulations with 
$k_g\rinit$ from $3.32$ to $4.52$ and the gas parameters
$(\Nres=128,\,k_g=1,\,f_g=0.01)$. The stacked curves are again well
aligned. We conclude that the soliton growth rate depends only on the
initial gas parameters and the instantaneous 
soliton mass (or radius),
and is insensitive to the previous evolution of the soliton-gas
system. This justifies combination of the results extracted from
different runs at different stages of simulations.

The above results suggest that the gas distribution remains close to
Maxwellian during the simulations with solitons. 
We have measured the distribution directly at different moments of
time and have seen that it is compatible with Maxwellian, though the
measurement is rather noisy, see \cref{fig:Maxwell} in
\cref{sec:levkov}. 
This is in stark contrast with simulations \cite{Levkov:2018kau}
without initial soliton where the gas distribution exhibits distinct
evolution on the time scale $\tau_{\rm rel}$ (\cref{trelax})
towards populating low-momentum modes which culminates in the soliton
formation. However, as discussed in \cref{sec:levkov}, the
distribution appears to return to Maxwellian after the soliton is
formed. We also 
find that the growth of the soliton mass, though
faster than in the Maxwellian gas right after the formation, approaches
the Maxwellian rate within time of order $\tau_{\rm rel}$, see
\cref{fig:stack_fig}. This gives another evidence that the presence of
the soliton ``Maxwellizes'' the gas. 

\begin{figure}[t]
     \centering
         \includegraphics[width=0.75\textwidth]{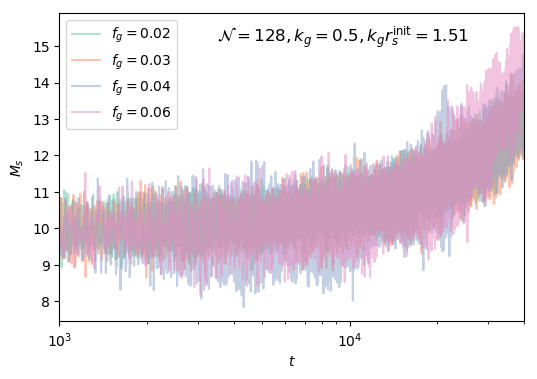}
\caption{Growth of the soliton mass in the simulations with the same
  values of
  $(\Nres=128,\,k_g=0.5,\, k_g\rinit=1.51)$ and varying $f_g$. The
  time axis in different runs has been scaled by $f_g^2$ and
  normalized to the case $f_g=0.06$. The time span of the curves is
  restricted to half of the relaxation time (\ref{trelax}) and covers
  the portion of the data used in the measurement of the soliton
  growth rate.
\label{fig:stack_scale}    
} 
\end{figure}

The analytic derivation of \cref{sec:theory} implies that at fixed
$k_gr_s$ the soliton growth/eva\-po\-ra\-tion rate is proportional to
$\rho_g^2/k_g^6\propto f_g^2$. To verify if this scaling holds in the
simulations, we perform several runs with the same ${\cal N}$, $k_g$
and $\rinit$, 
but different $f_g$. We measure the time dependence of the soliton
mass and scale the time axis by $f_g^2$. The results are shown in
\cref{fig:stack_scale}. We see a satisfactory agreement between
different curves. A slightly faster growth of the curve with the
highest value of $f_g$ at late times can be due to the fact that
the gas in this case is closer to the Jeans instability leading to the
development of an overdensity (proto-halo) around the soliton. We have
clearly seen this overdensity in the runs with the parameters near the
Jeans instability limit (\ref{eqn:jeans_instability}) and observed
that it is correlated with the increase of the ratio
$\Gamma_s/f_g^2$. The associated bias is comparable to the other
uncertainties in the measurement of $\Gamma_s$ and is included in
the error bars for our final results in the next section.

\subsection{Results}
\label{sec:results}

\begin{figure}[t]
\begin{center}
 \includegraphics[width=1.0\textwidth]{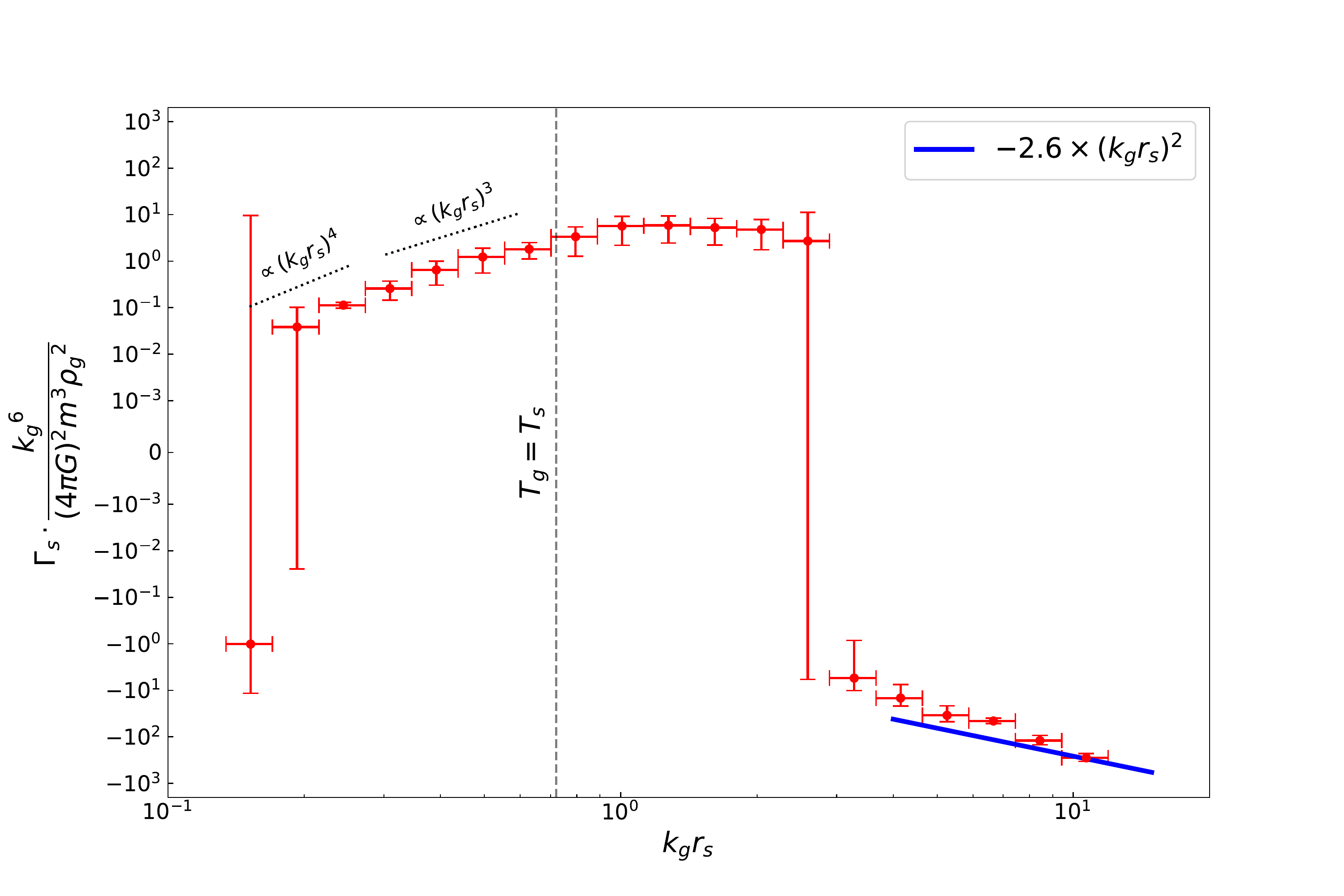}
\caption{
The soliton growth/evaporation rate as function of $k_gr_s$ --- the
product of the gas momentum and the soliton half-density radius. 
The cumulative dependence is constructed using 3900 data points
extracted from 195 independent simulations with different gas and
soliton parameters. 
The data are binned on logarithmic scale in
$k_gr_s$. Each dot gives the average value of the growth rate in the
bin, while the vertical error bars correspond to the standard deviation
within the bin. The blue solid line shows the asymptotic dependence
predicted by \cref{F2new}. At small $k_gr_s$ the dotted 
lines indicate possible power-law dependences. The
dashed vertical line marks the value of $k_gr_s$ corresponding to the
equality of the gas and soliton virial temperatures, $T_g/T_s= 1$.  
}
  \label{fig:growth}
  \end{center}
\end{figure}

In this section, we construct the cumulative dependence of $\Gamma_s$
on the soliton and gas parameters. As explained above, each simulation
run produces 20 data points $(r_s,\Gamma_s)$. We collect the data
points from 195 runs and bin them in logarithmic scale in $k_gr_s$. In
each bin we compute the average value and variance of 
\be
\Gamma_s\times \frac{(4\pi)^3}{ f_g^2}=\Gamma_s\times \frac{k_g^6}{\rho_g^2}\;. 
\ee
The results of this procedure are shown in \cref{fig:growth}. Note
that we restore the dimensionful constants in the scale of $\Gamma_s$
in the figure. 

Consistently with the analysis
of \cref{sec:theory}, the growth rate is positive at small $k_gr_s$
(heavy solitons) corresponding to the soliton growth, and is negative
at large $k_gr_s$ (light solitons) corresponding to
evaporation. Moreover, the data points with the largest values of
$k_gr_s$ match the asymptotic dependence (\ref{F2new}), including the
numerical coefficient (\ref{Clsnum}),\footnote{Recall the
  proportionality between $\nu$ and $k_g r_s$, \cref{eq:alpharatio}.}
\be
\label{Gammalast}
\Gamma_s\simeq -2.6\times \frac{(4\pi G)^2m^3\rho_g^2}{k_g^6}\,(k_gr_s)^2\;.
\ee
This dependence is shown by the blue line. Thus, we conclude that the
asymptotics (\ref{F2new}) are reached already at $k_gr_s\gtrsim
5$. The transition from evaporation to growth happens at $k_gr_s\sim
2.5$ which is in reasonable agreement with the naive estimate
(\ref{nucrit}). In terms of the gas and soliton virial temperatures,
it corresponds to $T_g/T_s\simeq 12$.

For lower $k_gr_s$ the soliton grows. The growth rate stays almost
constant in the range $0.7<k_gr_s<2$ where it is comparable to the
inverse of the gas
relaxation time $\tau_{\rm rel}^{-1}$, see \cref{trelax}. The lower
end of the plateau corresponds to the equality of the gas and soliton
virial 
temperatures, $T_g/T_s= 1$, which is marked by the dashed vertical 
line in \cref{fig:growth}. 

At $k_gr_s<0.7$ (equivalently $T_g/T_s<1$) the growth rate quickly decreases. We
find that this decrease is consistent with a power law
\be
\label{powerfit}
\Gamma_s\propto (k_gr_s)^n
\ee
with $n\simeq 3$ indicated by the dotted line in the
plot. The points with the smallest values of $k_gr_s$ hint at a
steepening dependence with $n=4$ at $k_gr_s\to 0$, in agreement
with the analytic estimate
(\ref{eq:heavyS_rate}).
There are, however, several caveats that prevent us from claiming
that we have reached the heavy soliton asymptotics. First,
as pointed out in \cref{sec:heavysoliton}, the expression
(\ref{eq:heavyS_rate}) has been obtained under the assumption that the
contribution of the bound states into the soliton growth scales with
$k_gr_s$ in the same way as the contribution of states from
continuum. This assumption must be verified by analyzing the kinetic
cascade in the soliton--gas system which is beyond the scope of the
present paper. Second, the low-$(k_gr_s)$ bins in our simulations are
at the extreme of the numerical resolution and close to the threshold
for halo formation. Therefore they can be affected by
systematics.
Without the three lowest-$(k_gr_s)$ bins the numerical
data are compatible with a shallower slope $n=2$. 
All in all, the heavy soliton limit is challenging both to numerical
and analytical methods. 
Taking into account the uncertainties, 
we conservatively conclude that the power $n$ in 
\cref{powerfit} for heavy solitons lies in the range
$2\leq n\leq 4$. More work is needed to pin
down the precise asymptotic value of $n$ at $k_gr_s\to 0$.

%==========================================================================
\section{Discussion and outlook}
\label{sec:conclusion}
%==========================================================================

\paragraph{Absence of kinetic equilibruium.}
We have found that a soliton (boson star)
immersed into a homogeneous Maxwellian axion gas evaporates if its virial
temperature is about 12 times lower than the virial temperature of the
gas, and grows otherwise. This rules out the possibility of a stable
kinetic equilibrium between the gas and the soliton.

\paragraph{Evaporation of light solitons.}
Though evaporation of cold solitons 
may at first sight appear surprising, the mechanism
behind it is quite intuitive. 
Being a self-gravitating system, the soliton possesses
negative heat capacity. Thus, a transfer of energy from the hot gas to
the cold soliton makes the latter even colder. This leads to a
run-away of the soliton temperature, and hence its mass, towards zero.

The parametric dependence of the evaporation rate can be estimated
using the following simple considerations.\footnote{We thank Neal Dalal
and Junwu Huang for the discussion on this topic.} Wave interference
in the axion gas produces density inhomogeneities with the
characteristic size of half de Broglie wavelength
$\lambda_a/2=\pi/k_g$. These inhomogeneities can be though of as
quasi-particles with the mass $M_{qp}\sim \rho_g(\lambda_a/2)^3$
\cite{Hui:2016ltb}. A single quasi-particle colliding with the soliton
transfers to it a recoil momentum 
\be
\delta p\sim \frac{G M_s M_{qp}}{r_s^2}\cdot \frac{r_s}{v_{qp}}\;,
\ee
where $v_{qp}\sim k_g/m$ is the quasi-particle velocity, and $r_s$
appears as the typical impact parameter. This implies the soliton
recoil energy
\be
\delta E_s\sim \frac{\delta p^2}{2M_s}\sim\frac{G^2M_{qp}^2 M_s}{2r_s^2
v_{qp}^2}\;.
\ee
Since the size of the quasi-particle is smaller than $r_s$ for the
light soliton, the recoil energy is distributed non-uniformly
throughout the soliton volume. This 
leads to excitation of its normal modes. The number of
axions that get excited from the ground state and hence get lost by
the soliton is of order $\delta N_s\sim -\delta E_s/|\E_s|$. Combining
everything together, we obtain the mass loss of the soliton in a
single quasi-particle collision,
\be
\frac{\delta M_s}{M_s}\sim -\frac{G^2 M_{qp}^2 m^2}{2v_{qp}^2}\;,
\ee
where we have used that $|\E_s|r_s^2\sim 1/m$. To obtain the
evaporation rate, we have to multiply this result by the number of
quasi-particles bombarding the soliton in a unit of time, 
$J_{qp}\sim 4\pi r_s^2 (\lambda_a/2)^{-3} v_{qp}$. In this way we
arrive at
\be
\Gamma_s\sim -\frac{2\pi^4G^2m^3\rho_g^2}{k_g^6}\,(k_g r_s)^2\;,
\ee
which agrees with the exact expression (\ref{Gammalast}) 
obtained from the kinetic theory, up to a factor about $0.5$.

We have seen that the threshold for evaporation is set by the equality
of the evaporation rate and the relaxation rate in the gas --- a
competing process leading to the soliton
formation~\cite{Levkov:2018kau}. This explains why the solitons that
are formed 
in the gas always have virial temperature comparable
to that of the gas: they are just hot (and heavy) enough to survive. 

In what physical situation can the soliton evaporation be relevant?
For fuzzy dark matter, this is the case when a small subhalo with low
velocity dispersion and light solitonic core falls into a bigger halo
with higher velocity dispersion. Evaporation then adds a new
destruction mechanism for the subhalo soliton, on top of the tidal
stripping \cite{Du:2018qor}. The time scale of evaporation is given by
the inverse of $|\Gamma_s|$, 
\begin{equation}
    t_{\rm evap} \simeq 2.4\times 10^{9} \,
\left( \frac{ m }{ 10^{-21} \, {\rm eV} }\right)^3\,
\left( \frac{ \rho_g }{ {0.3 \,{\rm GeV/ cm^3}} }\right)^{-2}\,
\left( \frac{v_g }{ 30 \, {\rm km/s} }\right)^6 \,
\left( \frac{k_g r_s }{ 10 }\right)^{-2} 
{\rm yr} \;,
\end{equation}
where $\rho_g$ and $v_g$ should be taken as the density and velocity
dispersion of the bigger halo at the orbit of the soliton. The
evaporation time is very sensitive to the halo parameters and can be
longer or shorter than the age of the universe depending on their
precise values. 
The evaporation should be also taken into account in the evolution of
boson stars in merging QCD axion miniclusters. Though here the
particle mass is much higher, the evaporation time can still be much
shorter than the age of the universe due to the very small velocity
dispersion $v_g\sim 10^{-5}\,{\rm km/s}$ in the miniclusters and their
extremely high density $\rho_g\gtrsim 10^6\,{\rm GeV/cm^3}$
\cite{Kolb:1994fi}. 

\paragraph{Growth of heavy solitons.} 
For solitons with virial temperature above the evaporation threshold
($T_s\gtrsim 0.1\,T_g$) we have found that the growth rate quickly
decreases once the soliton temperature exceeds that of the gas. This
result is in qualitative agreement with other works
\cite{Levkov:2018kau,Chen:2020cef}. The growth rate of heavy solitons
measured from our numerical simulations is consistent with the power
law (\ref{powerfit}) with $n$ between $2$ and $4$. We have presented
analytic arguments favoring $n=4$ in the limit $k_gr_s\to 0$, which
is compatible with the numerical data in the lowest
$k_gr_s$ bins. These bins, however, suffer from large uncertainties and
it remains unclear if the range $k_gr_s\gtrsim 0.2$
probed 
in the
simulations is sufficient to reach into the asymptotic heavy
soliton regime. 

The power-law dependence of the rate (\ref{powerfit})
translates into power-law growth of the soliton mass,\footnote{Recall
  that $r_s\propto M_s^{-1}$, whereupon the evolution equation for
  the mass is easily integrated.}
\be
\label{masspowerfit}
M_s\propto t^\alpha\;,~~~~~~\alpha=1/n\;.
\ee   
Ref.~\cite{Levkov:2018kau} established that $\alpha=1/2$ provides a
good fit to the soliton growth right after formation, whereas
Ref.~\cite{Chen:2020cef} found a dramatic flattening of the soliton
mass curve at
late times corresponding to $\alpha=1/8$. The results of
Ref.~\cite{Levkov:2018kau} are consistent with ours, though our
central value for the power $n=3$ predicts a somewhat shallower
dependence with $\alpha=1/3$. The steep growth observed in
\cite{Levkov:2018kau} might be due to a short duration of the
simulations. Indeed, by carrying out numerical experiments with the
same setup as in \cite{Levkov:2018kau} (see \cref{sec:levkov}) and
fitting the soliton mass with the formula (\ref{masspowerfit}), we
have 
observed a correlation of the best-fit index $\alpha$ 
with the soliton lifetime: $\alpha$ is
about $1/2$ for newly formed solitons and descreases down to $1/4$ for
grown-up solitons long after the relaxation time (see
\cref{fig:tend}). This trend is in agreement with our main simulations
where we see indications of increasing $n$, and hence decreasing
$\alpha$, for heavier solitons.
However, at this point the numerical data are rather inconclusive as to
the robustness of this trend and the asymptotic value of $\alpha$ at
$t\to\infty$. 

On the other hand, we do not see any evidence for the low 
$\alpha=1/8$ found in \cite{Chen:2020cef}. Moreover, our analytic
considerations suggest that the asymptotic value of $\alpha$ is at
least as high as $1/4$. The discrepancy may be due to the difference
in the setups. We study a soliton in a homogeneous gas, whereas
Ref.~\cite{Chen:2020cef} considers a soliton in the center of an axion
halo. It is conceivable that suppression of the soliton growth in the
latter case stems from its back reaction on the halo.
It will be interesting to explore this possibility in more
detail in future.  

\paragraph{Soliton-host halo relation.}
One can ask whether our results have any implications for the
soliton-host halo relation. The answer is: Not directly, because in the
cosmological setting the solitons were found to form during the initial
halo collapse when axions are not yet in the kinetic regime. Still,
with some degree of extrapolation, one can argue that our results make
unlikely formation of a light soliton since it would be evaporated by
the fast axions from the halo. This sets a lower bound on the soliton
mass which is just a factor of a few lower than $M_{s}^{\rm SSH}$,
the mass corresponding to the soliton-host halo
relation.\footnote{Note that by the soliton-host halo relation we
  understand here correlation between the soliton mass and the
  virial temperature of the halo, while in the literature the
  soliton-host halo relation is commonly formulated in terms of the
  halo mass. We believe that the former formulation reflects better
  the underlying physical mechanisms behind the relation.}
Heavier solitons can, in principle, form with arbitrary masses and
will continue growing upon the halo virialization. The time scale for
this growth can, however, be very long and exceed the age of the
universe when the soliton mass exceeds $M_{s}^{\rm SSH}$. Moreover, it
is natural to speculate that the solitons are more likely to form as
light as they can which singles out $M_{s}^{\rm SSH}$ as the sweet
spot. This reasoning still does not tell us how far the soliton-host
halo relation can be extrapolated in the parameter space. In
particular, we do not know whether the solitons form in any halo and
for any value of axion mass, or for some parameters their formation
becomes improbable. More work is needed to answer these questions.  

\paragraph{Persistence of Maxwell distribution.} It is known that
without a soliton the velocity distribution of axion gas relaxes
towards thermal form with high population of low-momentum modes
\cite{Levkov:2018kau}. We have found evidence that the presence of
soliton changes the picture. In this case the Maxwell distribution
appears to persist on timescales significantly longer than the kinetic
relaxation time. Moreover, in the simulations with soliton formation
we observed restoration of the Maxwell distribution after a transient
period with enhanced population of low-momentum modes preceding the
birth of the soliton. This ``Maxwellization'' manifests itself
indirectly in the universality of the soliton mass evolution in
simulations with different histories (figs.~\ref{fig:stack},
{\ref{fig:stack_fig}}), as well as in the directly measured momentum
distribution at different moments of time (\cref{fig:Maxwell}). The
latter, however, is subject to large temporal fluctuations which
presently do
not allow us to move beyond qualitative statements. It will be
interesting to study this phenomenon more quantitatively in future by
developing methods of measuring the momentum distribution with reduced
noise. A complementary approach would be to track the distribution of
axions in energy, instead of momentum, as suggested in
Ref.~\cite{Levkov:2018kau}.

%==========================================================================
\acknowledgments
%==========================================================================

We are grateful to Asimina Arvanitaki, 
Diego Blas, Kfir Blum, Tzihong Chiueh, Neal Dalal, Junwu Huang, 
Lam Hui, Dmitry Levkov, David Marsh, 
Jens Niemeyer, Alexander Panin, Hsi-Yu Schive, Bodo Schwabe, Pierre
Sikivie, 
Igor Tkachev, and Tak-Pong Woo 
for useful discussions. 
J.H.-H.C.~acknowledges the generosity of Eric and Wendy Schmidt by recommendation of the Schmidt Futures program.
The work of S.S. is supported by
the 
Natural Sciences and Engineering Research Council (NSERC) of Canada.
Research at Perimeter Institute is supported in part by the Government
of Canada through the Department of Innovation, Science and Economic
Development Canada and by the Province of Ontario through the Ministry
of Colleges and Universities. 
W.X. is supported in part by the U.S. Department of Energy 
under grant DE-SC0022148 at the University of Florida.

\appendix

\section{Classical derivation of the soliton growth rate}
\label{app:class}

In this appendix we derive the expression (\ref{eq:NsGen}) for the
soliton growth rate as the consequence of the classical equations of
motion. It is convenient to integrate out the gravitational potential
and rewrite the Schr\"odinger--Poisson system as a single equation
with non-local interaction,
\be
\label{SPsingle}
i\d_t\psi+\frac{\Delta \psi}{2m} -4\pi Gm^2\psi\frac{1}{\Delta}|\psi|^2=0\;,
\ee
where $\tfrac{1}{\Delta}$ denotes the Green's function of the
Laplacian. Clearly, this equation conserves the total mass of axions
in a box $M_{\rm tot}=m\int d^3x |\psi|^2$. Now, we make the split
(\ref{split}) into the soliton and gas and, using the fact that the
soliton is a solution of \cref{SPsingle}, obtain the equation
for the gas component,
\be
\label{gaseqapp}
\begin{split}
i\d_t\psi_g+\frac{\Delta\psi_g}{2m}&-4\pi Gm^2\bigg[
\psi_g\frac{1}{\Delta}|\psi_s|^2
+\psi_s\frac{1}{\Delta}(\psi_s^*\psi_g)
+\psi_s\frac{1}{\Delta}(\psi_s\psi_g^*)\bigg]\\
&-4\pi Gm^2\bigg[
\psi_g\frac{1}{\Delta}(\psi_s^*\psi_g)
+\psi_g\frac{1}{\Delta}(\psi_s\psi_g^*)
+\psi_s\frac{1}{\Delta}|\psi_g|^2
+\psi_g\frac{1}{\Delta}|\psi_g|^2\bigg]=0\;.
\end{split}
\ee
In the first line we have grouped the terms that affect the gas field
at linear order, whereas the second line contains interactions. Note
that, despite the presence of the small factor $4\pi Gm^2$, all terms
in the first line are of the same order because $\psi_s$ is
proportional to $(4\pi Gm^2)^{-1/2}$, see
\cref{solitonWF}. Correspondingly, the leading interaction terms are
of order $\sqrt{4\pi G m^2}$. 

The mass of the gas is not constant. From \cref{gaseqapp} we have,
\be
\frac{dM_g}{dt}=m\frac{d}{dt}\int d^3 x|\psi_g|^2
=-(8\pi Gm^3) \Im\int
d^3x\bigg[
\psi_s^*\psi_g\frac{1}{\Delta}(\psi_s^*\psi_g)
+\psi_s^*\psi_g\frac{1}{\Delta}|\psi_g|^2\bigg]\;,
\ee
where we have canceled the boundary terms assuming periodic boundary
conditions. Since the total mass is conserved, this must be
compensated by the change in the soliton mass. Thus, we obtain for the
soliton growth rate,
\be
\label{Gammaclass}
\Gamma_s=\frac{8\pi Gm^3}{M_s}
 \Im\int
d^3x\bigg[
\psi_s^*\psi_g\frac{1}{\Delta}(\psi_s^*\psi_g)
+\psi_s^*\psi_g\frac{1}{\Delta}|\psi_g|^2\bigg]\;.
\ee

If we neglect the interaction terms in \cref{gaseqapp}, it admits a
set of periodic-in-time solutions. We decompose the gas field in
these eigenmodes,\footnote{Due to the last term in the first line of
  \cref{gaseqapp} that mixes $\psi_g$ and $\psi^*_g$, the eigenmodes
  contain both positive and negative frequencies \cite{soliton2}. To
  avoid cumbersome expressions, we neglect this subtlety in the following
  discussion. It does not affect the final result for the soliton
  growth rate.}
\be
\label{psigdecomp}
\psi_g(t,\x)=\sum_i a_i(t)\e^{-i\E_i t}\psi_i(\x)\;, 
\ee
where the amplitudes $a_i(t)$ slowly vary due to the
interactions. Substituting into \cref{Gammaclass} we obtain,
\be
\label{Gammaclass1}
\Gamma_s=-\frac{2m}{M_s}\Im\bigg[
\sum_{i,j} a_i a_j\e^{-i(\E_i+\E_j-2\E_s)t}A'_{is,js}
+\sum_{i,j,k} a_i a_j a_k^*\e^{-i(\E_i+\E_j-\E_k-\E_s)t}A'_{is,jk}\bigg]\;,
\ee
where the scattering amplitude $A'_{is,jk}$ is defined in
\cref{Astrip}, and $A'_{is,js}$ is defined similarly with the $k$th
wavefunction replaced by the soliton. All terms in the first sum
quickly oscillate since the gas states are separated from the ground
state by an energy gap of order $|\E_s|$. Thus, they disappear once we
average the growth rate over time scales of order $|\E_s|^{-1}$ and
we omit them in what follows.

The second sum does not vanish upon time averaging because the
combination of energies in the exponent can be small. However, to
obtain the physical growth rate we also have to average
over random initial phases of the gas amplitudes. In the
absence of interactions the amplitudes $a_i$ in \cref{Gammaclass1}
coincide with the initial amplitudes $a_i^{(0)}$ and thus averaging
over their phases will give $\Gamma_s=0$. To obtain a non-zero result,
we have to take into account gas interactions. 

The first correction to the free gas field is due to
terms of order $\sqrt{4\pi Gm^2}$ in \cref{gaseqapp}. We can write it
schematically as
\be
\label{psig1}
\psi_g^{(1)}=(4\pi Gm^2)\,{\cal G}_{\rm ret}*\bigg\{
\psi_g^{(0)}\frac{1}{\Delta}\big(\psi_s^*\psi_g^{(0)}\big)
+\psi_g^{(0)}\frac{1}{\Delta}\big(\psi_s{\psi_g^{(0)}}^*\big)
+\psi_s\frac{1}{\Delta}|\psi_g^{(0)}|^2\bigg\}\;,
\ee
where $\psi_g^{(0)}$ is the free gas field and
${\cal G}_{\rm ret}$ is the retarded Green's function of the
operator in the first line of (\ref{gaseqapp}). Using the complete set
of eigenmodes, it can be written as,\footnote{For simplicity, we again
neglect the subtleties associated with the negative-frequency components
of the eigenmodes \cite{soliton2}.}
\be
\label{Gret}
{\cal G}_{\rm ret}(t-t',\x,\x')=\sum_i\int\frac{d\E}{2\pi}
\,\frac{\psi_i(\x)\psi_i^*(\x')}{\E-\E_i+i\epsilon}\,\e^{-i\E(t-t')}\;. 
\ee
Substituting this expression into (\ref{psig1}) and expanding
$\psi_g^{(1)}$ and $\psi_g^{(0)}$ into eigenmodes, we obtain the
first-order correction to the amplitudes,
\be
\begin{split}
a_i^{(1)}=-\sum_{j,k}\bigg[&
a_j^{(0)}a_k^{(0)}
\frac{\e^{-i(\E_j+\E_k-\E_i-\E_s)t}}{\E_j+\E_k-\E_i-\E_s+i\epsilon}
A'_{ks,ji}\\
&+a_j^{(0)}{a_k^{(0)}}^*
\frac{\e^{-i(\E_j-\E_k-\E_i+\E_s)t}}{\E_j-\E_k-\E_i+\E_s+i\epsilon}
(A'^*_{ks,ij}+A'^*_{is,kj})\bigg]\;.
\end{split}
\ee
Next, we insert this expression into the first-order contribution to
the soliton growth rate,
\be
\Gamma_s^{(1)}=-\frac{2m}{M_s}\sum_{i,j,k}
\Big(a_i^{(1)}a_j^{(0)}{a_k^{(0)}}^*
+a_i^{(0)}a_j^{(1)}{a_k^{(0)}}^*
+a_i^{(0)}a_j^{(0)}{a_k^{(1)}}^*\Big)
\e^{-i(\E_i+\E_j-\E_k-\E_s)t}A'_{is,jk}\;,
\ee 
and average over the phases of $a_i^{(0)}$ using
\be
\langle a_i^{(0)}a_j^{(0)}{a_{i'}^{(0)}}^*{a_{j'}^{(0)}}^*\rangle
=f_if_j(\delta_{ii'}\delta_{jj'}+\delta_{ij'}\delta_{ji'})\;.
\ee
Upon a somewhat lengthy, but straightforward calculation, we arrive at
\be
\begin{split}
\langle \Gamma_s\rangle=\frac{m}{M_s}
\Im\sum_{i,j,k}\bigg\{&\frac{f_jf_k+f_if_k}{\E_k-\E_j-\E_i+\E_s+i\epsilon}
|A'_{is,jk}+A'_{js,ik}|^2\\
&+\frac{f_jf_k}{-\E_i+\E_s+i\epsilon}
\Big[(A'_{is,jj}+A'_{js,ij})(A'^*_{is,kk}+A'^*_{ks,ik})+{\rm
  h.c.}\Big]\\
&+\frac{f_if_j}{\E_i+\E_j-\E_k-\E_s-i\epsilon}|A'_{is,jk}+A'_{js,ik}|^2\bigg\}.
\end{split}
\ee
In the final step we use the formula
\be
\Im\frac{1}{z+i\epsilon}=-i\pi\delta(z)\;.
\ee
Then the second term vanishes because $\E_i\neq \E_s$, whereas the
rest of the terms reproduce \cref{eq:NsGen}. Thus, we have shown that
the classical derivation leads to the same soliton
growth rate as the quantum mechanical one, upon averaging over the
ensemble of gas realizations with different initial phases. 

The above derivation also allows us to estimate the r.m.s. fluctuations
of $\Gamma_s$ in individual realizations. To this aim, let us return
to \cref{Gammaclass1} and smooth it with a Gaussian filter over time
scales $\langle\Gamma_s\rangle^{-1}>\tau\gg|\E_s|^{-1}$. We obtain,
\be
\Gamma_s=-\frac{2m}{M_s}\Im\sum_{i,j,k} a_ia_ja_k^*
\e^{-i(\E_i+\E_j-\E_k-\E_s)t}A'_{is,jk}\e^{-\tau^2(\E_i+\E_j-\E_k-\E_s)^2/2}\;.
\ee 
To get the r.m.s. fluctuations, we subtract $\langle\Gamma_s\rangle$,
square the result and average over the gas phases. In the latter step
we can replace $a_i$ with $a_i^{(0)}$ to obtain the leading
contribution. Retaining only the unsuppressed terms we obtain,
\be
\begin{split}
\langle\delta\Gamma_s^2\rangle&\simeq
\bigg(\frac{m}{M_s}\bigg)^2\sum_{i,j,k}f_if_jf_k\,
|A'_{is,jk}+A'_{js,ik}|^2\e^{-\tau^2(\E_i+\E_j-\E_k-\E_s)^2}\\
&\simeq\frac{\sqrt{\pi}}{\tau}
\bigg(\frac{m}{M_s}\bigg)^2\sum_{i,j,k}f_if_jf_k\,
|A'_{is,jk}+A'_{js,ik}|^2\delta(\E_i+\E_j-\E_k-\E_s)\;.
\end{split}
\ee 
Comparing this with the expression (\ref{eq:NsGen}) for the rate, 
we get an estimate
\be
\langle\delta\Gamma_s^2\rangle\sim\frac{1}{\tau}\frac{m}{M_s} f_g
\langle\Gamma_s\rangle\;.
\ee
The fluctuations are much smaller than the average if 
$\langle\Gamma_s\rangle\tau\gg mf_g/M_s$ which can be always achieved
by an appropriate choice of the smoothing scale, as long as the number
of particles in the soliton is much larger than the occupation numbers
of individual modes in the gas, $M_s/m\gg f_g$.

\section{Formation of axion soliton from the gas}
\label{sec:levkov}

\begin{figure}[t]
\begin{center}
 \includegraphics[width=0.7\textwidth]{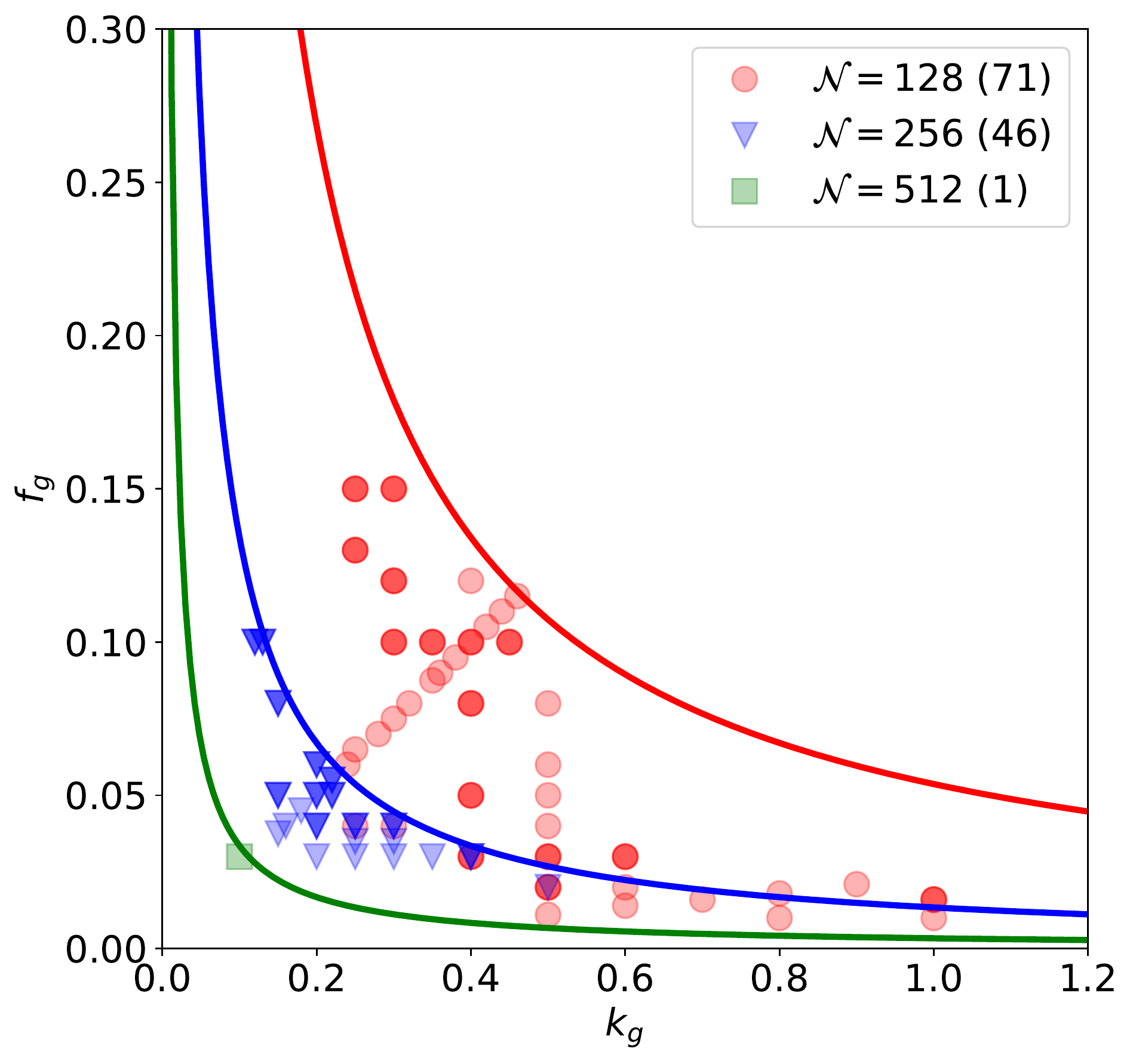}
\caption{ 
Gas parameters for the simulations with soliton
  formation. Solid lines bound the regions without Jeans instability
  for different simulation box sizes (see \cref{eqn:jeans_instability}). 
 The number of runs on different lattices is indicated in 
  parenthesis.
  \label{fig:formation_input}
}
  \end{center}
\end{figure}

In this appendix we report the results of simulations with 
formation of the soliton from the gas. We use the same numerical
scheme and initial conditions for the gas as described in
\cref{sec:setup}, but we do not put the initial soliton. Instead, we
wait for the soliton to emerge spontaneously.
The purpose
of these simulations is twofold. First, we cross-check our numerical
approach by comparing with the simulations carried out in
\cite{Levkov:2018kau}.\footnote{We thank Dmitry Levkov and Alexander
  Panin for sharing with us their results for a detailed comparison.}  
Second, we investigate to what extent the evolution of spontaneously
formed solitons is similar to the evolution of the solitons inserted
into the gas from the start.

We perform 118 independent simulations with the parameters summarized
in \cref{fig:formation_input}. The parameter space is restricted by
the requirement of absence of the Jeans instability, so that the gas
does not form a halo and remains homogeneous. 

\begin{figure}[t]
\begin{center}
 \includegraphics[width=0.9\textwidth]{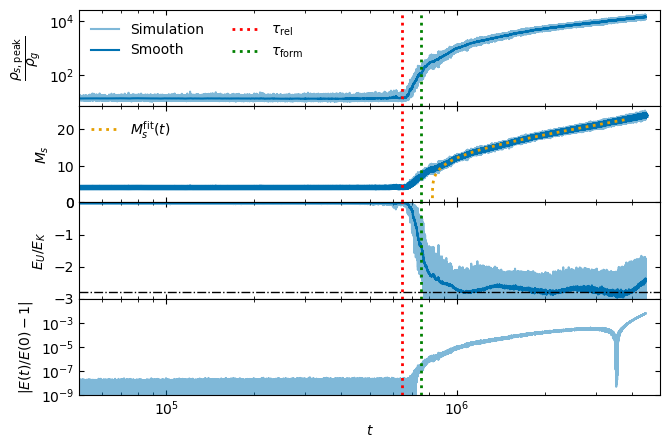}
\caption{
Example of spontaneous soliton formation in axion gas with parameters
${(\Nres=128,k_g=0.5,f_g=0.02)}$. From top to
bottom: maximal density in the simulation box, 
soliton mass estimated from the peak density,
virial ratio $E_U/E_K$, total energy in the box. Thick blue lines show
the smoothed dependences. Yellow dotted line is the fit
(\ref{eq:fitting_fig}). Vertical red and green dotted lines mark the
relaxation time (\ref{trelax}) and the measured soliton formation
time, respectively.
  \label{fig:formation_case}
}
  \end{center}
\end{figure}

Figure \ref{fig:formation_case} shows the results of a typical
simulation run. The maximal axion density within the simulation box 
remains small for time
less than the relaxation time (\ref{trelax}) marked with the red
dotted line. Then it starts growing which signals the formation of a
soliton. As in \cref{sec:simulation}, we determine the
soliton mass from its peak density using \cref{eq:Mg}. We also
construct smoothed peak density and soliton mass using a top-hat
filter with the width $t_{\rm width}=70/k_g^2$.  The smoothed
dependences are shown in the figure with thick blue lines.

To pin down the moment of soliton formation, we use the method
proposed in \cite{Veltmaat:2018}. We identify the density maximum
within the simulation box and compute the kinetic ($E_K$) and
potential ($E_U$) energy in a spherical region around it. 
The radius of the sphere is chosen as the radius at which the
shell-averaged density drops to half of its peak value. To calculate
the kinetic energy, we evaluate the field gradient, 
subtract the center-of-mass velocity
contribution, square the result and integrate
over the volume of the sphere. The potential energy is approximated
by the potential energy of a uniform ball with the mass enclosed
inside the sphere. 
For a random peak in the gas the ratio $E_{U}/E_K$ is
close to zero, whereas for the soliton it obeys the virial
condition\footnote{The ratio is different from $-2$ because we
  consider only the inner part of the whole soliton.}  
 $E_{U}/E_K\simeq -2.8$. In \cref{fig:formation_case} we see that this
 ratio changes abruptly from $0$ to $-2.8$ around $t\sim\tau_{\rm
   rel}$. We identify the soliton formation time $\tau_{\rm form}$ as
 the moment when the smoothed curve $E_{U}/E_K$ crosses half of its
 virial value,
\be
E_U/E_K\Big|_{\tau_{\rm form}}=-1.4\;.
\ee
This time is marked with the green dotted line in the plot. We see
that it agrees well with the relaxation time $\tau_{\rm rel}$.

Ref.~\cite{Levkov:2018kau} suggested that upon formation the growth of
the soliton is described by a power-law 
\begin{equation}
M_s(t) = M_0\left( \dfrac{t}{\tau_0}-1\right)^\alpha
\label{eq:fitting_fig}
\end{equation}
with $\alpha=1/2$, $\tau_0=\tau_{\rm rel}$ and $M_0\simeq 12\pi
k_g$. To verify if this law is obeyed in our simulations, we fit the
smoothed soliton mass at $t>\tau_{\rm form}$ 
with the formula (\ref{eq:fitting_fig}) allowing
$\alpha$, $\tau_0$, $M_0$ to vary as free fitting parameters. The
fitting time range is restricted by the condition that the 
energy error 
$\left|E(t)/E(0)-1\right|$ does not exceed
$0.1\%$. 
The result of the
fit is shown by yellow dotted line 
in \cref{fig:formation_case}. The best-fit parameters for this run are
$\alpha=0.22$, $\tau_0=8.2\times 10^5$, $M_0=17.03$. 
Note that
the value of $\alpha$ is significantly lower than $1/2$. 
We will discuss shortly how this result can be reconciled with those
of Ref.~\cite{Levkov:2018kau}.

\begin{figure}[t]
\begin{center}
 \includegraphics[width=1.0\textwidth]{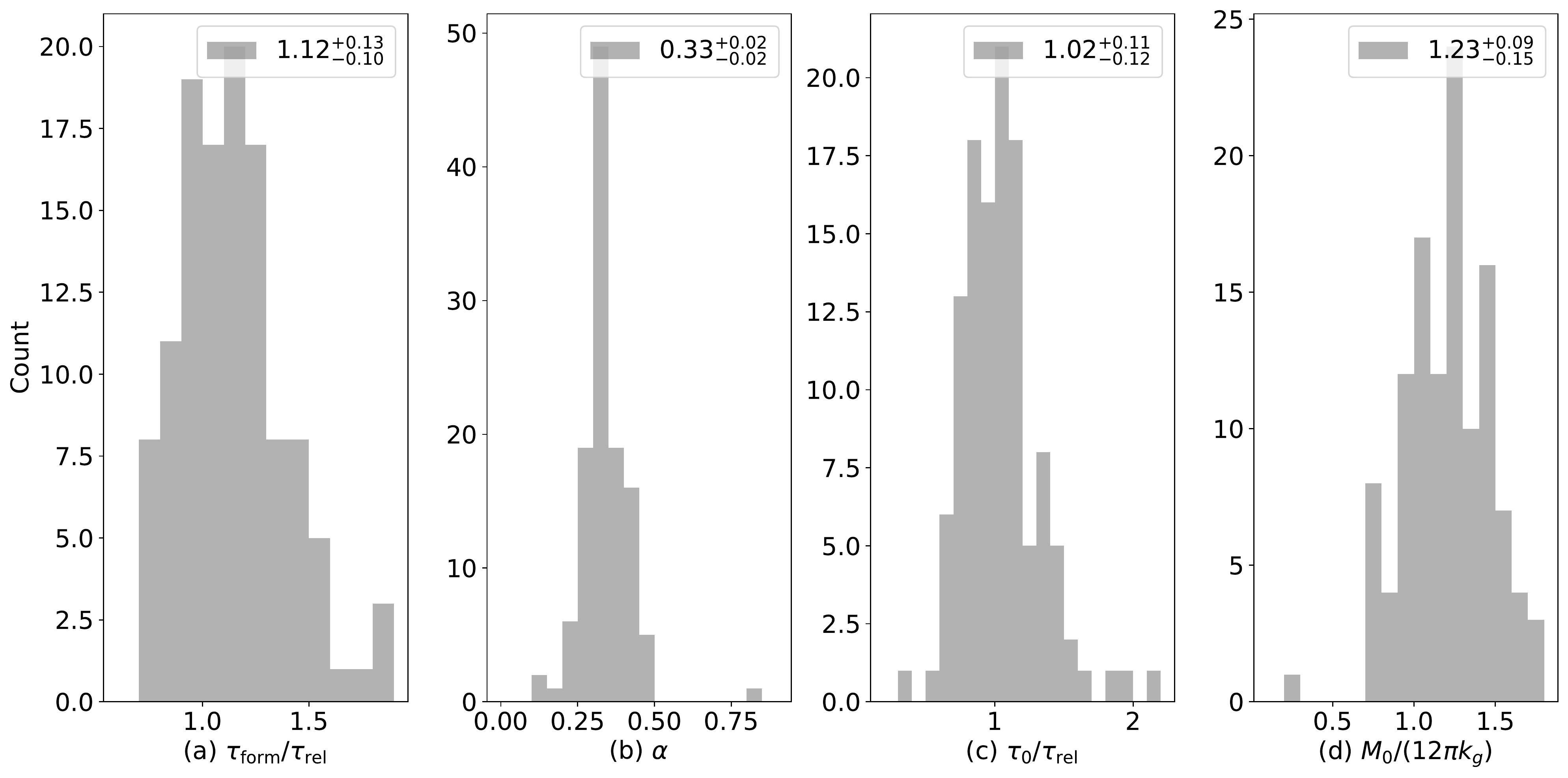}
\caption{ 
Results of the measurements in the simulations with soliton
formation. The histograms show the distributions 
of the soliton formation time $\tau_{\rm form}$, and the
parameters in the power-law fit (\ref{eq:fitting_fig}) 
of the soliton mass growth: $\alpha$, $\tau_0$, $M_0$. The
relaxation time $\tau_{\rm rel}$ is given by \cref{trelax} and $k_g$
is the gas momentum.
\label{fig:formation}
}
  \end{center}
\end{figure}

We repeat the above analysis for each of 118 runs and construct the
histograms of 
$\tau_{\rm form}$, $\alpha$, $\tau_0$, $M_0$ 
measured in different runs. These histograms are shown in  
\cref{fig:formation} together with their means and standard
deviations. The mean values of $\tau_{\rm form}$, $\tau_0$ and
$M_0$ are in good agreement with the findings of
\cite{Levkov:2018kau}. On the other hand, for the exponent we obtain a
lower mean, $\alpha=0.33\pm 0.02$. It is important to notice,
however, that the distribution of $\alpha$ is quite broad, extending
from\footnote{There are three outlier runs with very high
  ($\alpha\simeq 0.8$) and very low ($\alpha\simeq 0.1$)
  exponents. The origin of these large fluctuations is unknown.}  
$0.2$ to $0.5$. From the analysis in the main text we know
that the soliton growth rate decreases when the soliton gets
heavier. This suggests that the spread in $\alpha$ can arise due to
different soliton masses achieved in different simulations. In this
picture, the runs with larger duration should yield lower values of
$\alpha$ since the solitons in them have more time to grow. 

\begin{figure}[t]
\begin{center}
 \includegraphics[width=0.7\textwidth]{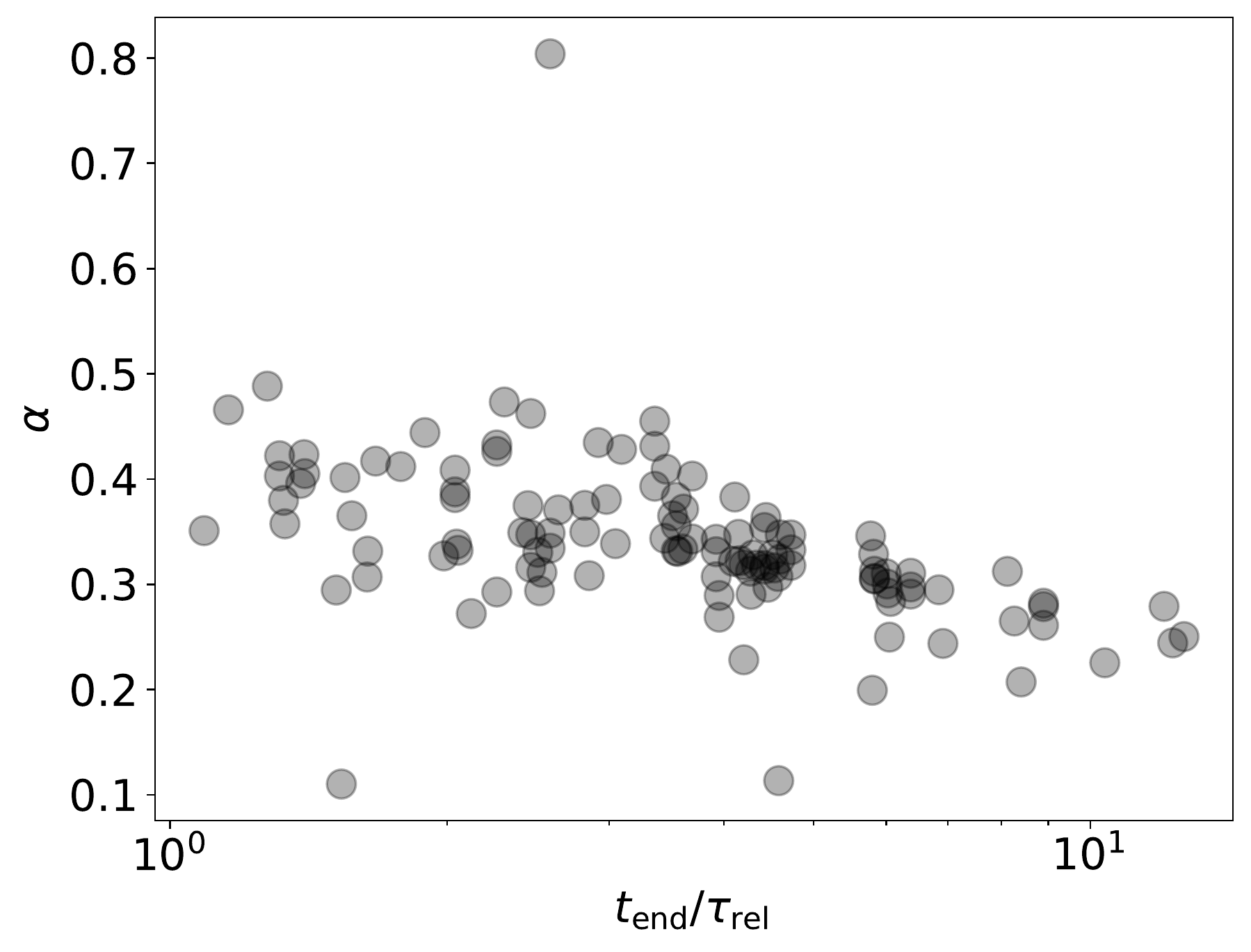}
\caption{The exponent in the power-law fit (\ref{eq:fitting_fig}) for
  the soliton mass against final simulation time $\tend$ measured in
  units of the relaxation time (\ref{trelax}).
 Longer simulations produce more massive solitons which have slower
 growth rate and hence lower values of $\alpha$.  
Three outlier
 simulations with $\alpha\simeq 0.8$ and $\alpha\simeq 0.1$ represent
 large fluctuations of unknown origin. 
  \label{fig:tend}
}
  \end{center}
\end{figure}

To check this expectation, we plot in \cref{fig:tend} the best-fit value
of $\alpha$ as function of the duration of the
simulation\footnote{More precisely, we take $t_{\rm end}$ to be the
  end of the time range used in the fit (\ref{eq:fitting_fig}).} in units of
relaxation time.
Apart from a few outliers, the bulk of the data exhibit a pronounced
anti-correlation between $\alpha$ and $t_{\rm end}/\tau_{\rm
  rel}$. The exponent varies from $\alpha\simeq 0.5$ for newly-born
solitons down to $\alpha\lesssim 0.25$ for long-lived solitons. Thus,
the value $\alpha=1/2$ found in \cite{Levkov:2018kau} can be explained
by short duration of the simulations used in the analysis, whereas
longer simulations carried out in the present work uncover a trend for
the decrease of $\alpha$ with time. This trend is consistent, both
qualitatively and quantitatively, with the results on heavy soliton
growth from the main text. Indeed, the scaling (\ref{powerfit}) of the
soliton growth rate implies
\be
\frac{1}{M_s}\frac{dM_s}{dt}\propto\frac{1}{M_s^n}~~~~~
\Longrightarrow~~~~~M_s\propto \bigg(\frac{t}{\tau_0}-1\bigg)^{1/n}\;,
\ee
which leads to the identification $\alpha=1/n$. Thus, the slow-down of
the soliton growth with $\alpha$ decreasing from $1/2$ to $1/4$ as
time goes on matches the steepening of the $\Gamma_s$ dependence on
$k_gr_s$ with $n$ increasing from $2$ to $4$ at smaller $k_gr_s$ (see
\cref{sec:results}). 

\begin{figure}[t]
\begin{center}
 \includegraphics[width=0.7\textwidth]{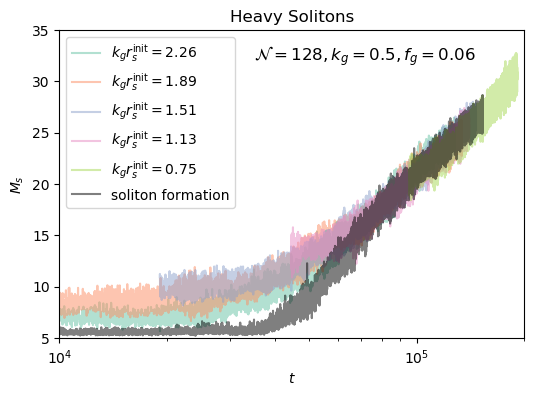}
\caption{Same as upper panel in 
\cref{fig:stack} with the addition of the soliton mass evolution 
from a run with
soliton formation (in grey). The spontaneously 
formed soliton approaches the same
growth rate as the solitons embedded in the gas from the start. 
  \label{fig:stack_fig}
}
  \end{center}
\end{figure}

The above match is non-trivial. The simulations of
\cref{sec:simulation} 
are
performed with Maxwellian gas and the growth rate is extracted from
time ranges shorter than half of the relaxation time to avoid any
significant change in the gas distribution. On the other hand, the
simulations in this appendix, by construction, span more than the
relaxation time. Moreover, it is known \cite{Levkov:2018kau} that the
soliton formation is preceded by a dramatic change in the gas
distribution with enhanced population of low-momentum modes. Thus, the
solitons in the two simulation suits are embedded in environments with
very different histories and their growth rate need not be the
same. Nevertheless, it turns out that the soliton growth exhibits a
remarkable universality. In \cref{fig:stack_fig} we superimpose the
time-dependent mass of a soliton born in the gas on top of the soliton
masses from out main simulation suit with solitons incorporated in
the initial conditions. We see that after a brief transient period of
a faster growth, the formed soliton approaches the same time
dependence as the solitons with the same mass 
that are present in the gas from the
start.

\begin{figure}[t]
\begin{center}
 \includegraphics[width=\textwidth]{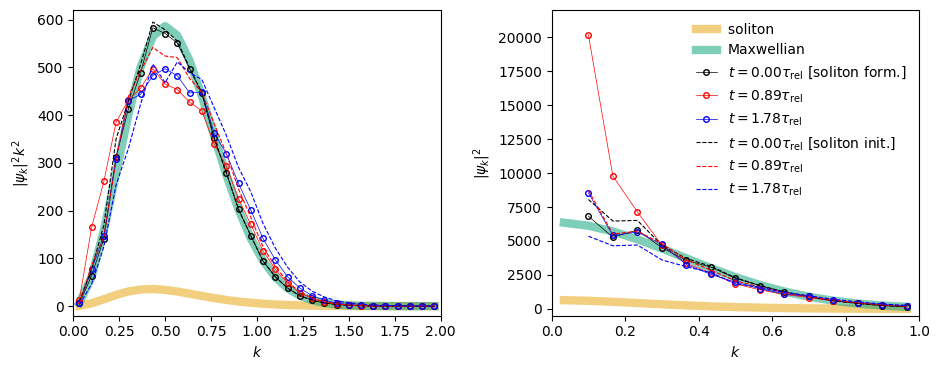}
\caption{
{\it Left panel:} Evolution of momentum distribution of axions in the
simulation box. 
The mode amplitudes are spherically
averaged over shells with fixed $k=|\k|$. 
{\it Right panel:} Zoom-in on the low-$k$ part
of the spectrum, where we divide the distribution by $k^2$ to make
the difference between curves more pronounced. The
distribution in a simulation with spontaneous formation of
the soliton from the gas ($\Nres=128,\,k_g=0.5,\,f_g=0.06$)
is shown by solid lines with circles. It is
sampled at three moments of time: at the beginning of the simulation
(black),
at the time before soliton formation formation (red) and after
the soliton has formed (blue). Just before the soliton
forms the distribution features a pronounced bump at low momenta which
disappears afterwards. For comparison, we show with dashed lines 
the distribution in a
simulation with soliton inserted in the initial conditions
($k_g\rinit=1.51$) sampled at the same time intervals. Maxwell
distribution corresponding to the input gas parameters
is shown with thick green line.  
The momentum wavefunction of the soliton with the mass achieved at
latest sampling point is plotted by thick yellow line.    
  \label{fig:Maxwell}
}
  \end{center}
\end{figure}

This suggests that the gas distribution restores its Maxwellian form
after the soliton formation. We check this conjecture by measuring the
amplitudes of axion modes $|\psi_\k|^2$ in the simulation from
\cref{fig:stack_fig} at several moments of time: at the beginning of
the simulation ($t=0$), before the soliton formation ($t=0.89\,\tau_{\rm
rel}$), and after the soliton has formed ($t=1.78\,\tau_{\rm rel}$). The
amplitudes are averaged over spherical shells with fixed values of
$k=|\k|$. The results are shown in \cref{fig:Maxwell} (solid
lines with circles). We see that right 
before the soliton formation, the distribution
develops a pronounced bump in the low-$k$ part of the spectrum,
consistently with the results of \cite{Levkov:2018kau}. This bump,
however, disappears after the soliton is formed and at late times the
distribution qualitatively resembles Maxwellian (shown by the thick green
line). We also superimpose in the same figure the distribution for the
run with soliton initially present in the gas sampled at the same
intervals from the start of the simulation
(dashed lines). The parameters of this run are
$(\Nres=128,\,k_g=0.5,\,f_g=0.06,\,k_g\rinit=1.51)$ and correspond to
the blue curve in \cref{fig:stack_fig}. In this case we see that
the distribution preserves the Maxwellian shape at all times, without
any excess at low-$k$ modes. We conclude that the presence of
the soliton affects the axion gas in a curious way: it stabilizes
the Maxwell distribution of axion momenta.

It is worth stressing that we are talking about the distribution in
the gas and not in the soliton itself. Though our numerical procedure
does not allow us to separate the two, we can compare the total
distribution to the wavefunction of the soliton in momentum
space. This is shown by thick yellow line in \cref{fig:Maxwell}. We take
the soliton mass to be $M_s=20$ corresponding to the latest sampling time. 
We see that the contamination of the distribution 
from the soliton is negligible. 

We do not attempt to explore this ``Maxwellization'' phenomenon further
in this work. The axion momentum distribution is
subject to significant temporal fluctuations which form an obstruction
for moving beyond qualitative statements. For a quantitative study,
one needs to devise less noisy probes. 
We leave this task for
future.

\section{Details of the numerical simulations}
\label{sec:nums}
\subsection{Convergence tests}
\label{sec:resolution}
In this work, we adopt second order drift-kick-drift 
operator (\ref{DKD}) to 
evolve wave function for each time step $\dt$.
The gravitational potential $\Phi$ and kinetic energy operators
$\Delta$ are calculated with CUDA Fast Fourier Transform
(cuFFT)\footnote{\url{https://developer.nvidia.com/cufft}}. 
We notice that the single precision of cuFFT causes $\approx 10\%$
mass loss in $10^6$ time steps. 
We therefore conduct the simulations in this work using the double precision. 
This makes the mass loss negligible (less than $10^{-6}$).

\begin{figure}[t]
\begin{center}
\includegraphics[width=0.5\textwidth]{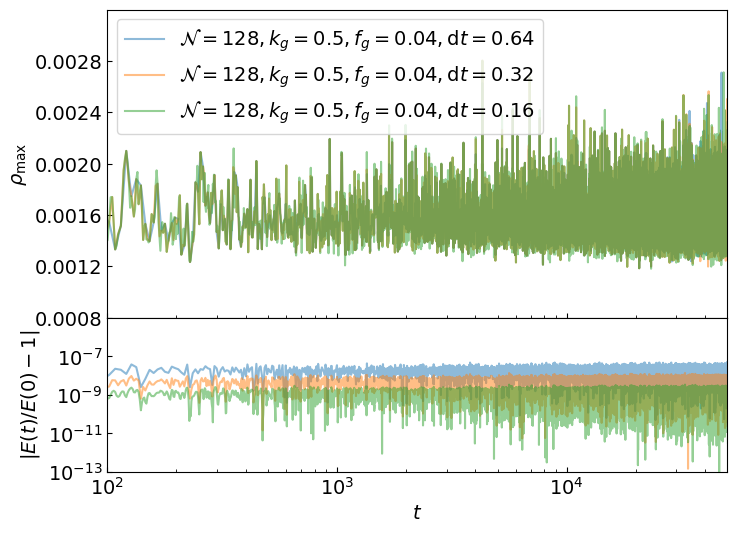}~~~~
 \includegraphics[width=0.49\textwidth]{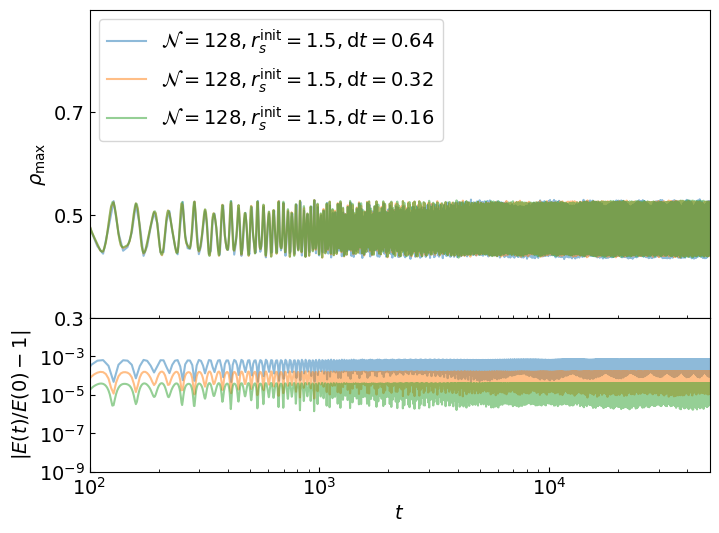}
\caption{Convergence tests in the simulations with pure gas (left) and
  an isolated soliton (right). In each case we perform three runs:
  one with the fiducial time step ${\rm d}t=0.64$, 
and two with time steps
  reduced by a factor of 2 and 4. The gas momentum is $k_g=0.5$,
  whereas the soliton radius is $\rinit=1.5$. The lattice size if
  $\Nres=128$ in both cases.
\label{fig:res_gas_sol}
 }
  \end{center}
\end{figure}

The requirement that the gas and the soliton must be resolved by the
spatial lattice puts and upper bound on the gas momentum
$k_g$ and a lower bound on the initial soliton size
$r_s^{\rm init}$ accessible in the simulations. 
To determine the domain of validity of our code, we perform
several convergence tests. First, we evolve the gas without the
soliton using three different time steps:  ${\rm d}t=2/\pi\simeq 0.64$
(our fiducial value), 
${\rm d}t=1/\pi\simeq 0.32$ and ${\rm d}t=1/(2\pi)\simeq
0.16$. 
The gas parameters in all three runs are
$(\Nres=128,\,k_g=0.5,\,f_g=0.04)$. The maximal density within the box
and the total energy measured in these runs are shown in the left panel
of \cref{fig:res_gas_sol}. We observe that the density curves
essentially coincide, while the energy error is proportional to $({\rm
  d}t)^2$, as it should. For our fiducial value of ${\rm d}t=2/\pi$, the
error 
stays well
below $10^{-7}$. We conclude that the gas with $k_g= 0.5$ is
comfortably resolved in our simulations.

Next, we repeat the same convergence test with an isolated soliton of
radius $\rinit=1.5$. The results are shown in the right panel of
\cref{fig:res_gas_sol}. Since the analytical template (\ref{eq:rhos})
used in the simulations to set the initial conditions slightly
deviates from the exact soliton profile, the soliton is initiated in
an excited state which leads to the oscillations of the peak
density. The oscillations obtained with three different time steps
match almost identically. The energy error also exhibits the proper
scaling, $|E(t)/E(0)-1|\propto ({\rm d}t)^2$. However, now it is
significantly larger, reaching up to $10^{-3}$ for the fiducial
${\rm d}t$. This is likely 
due to high frequency of the soliton phase rotation
$|\E_s|\simeq 0.52$ which is less resolved with the large time
step. Therefore, to correctly capture the evolution of the soliton
wavefunction, we restrict our simulations to $\rinit\geq 1.5$.

\begin{figure}[t]
\begin{center}
 \includegraphics[width=1.0\textwidth]{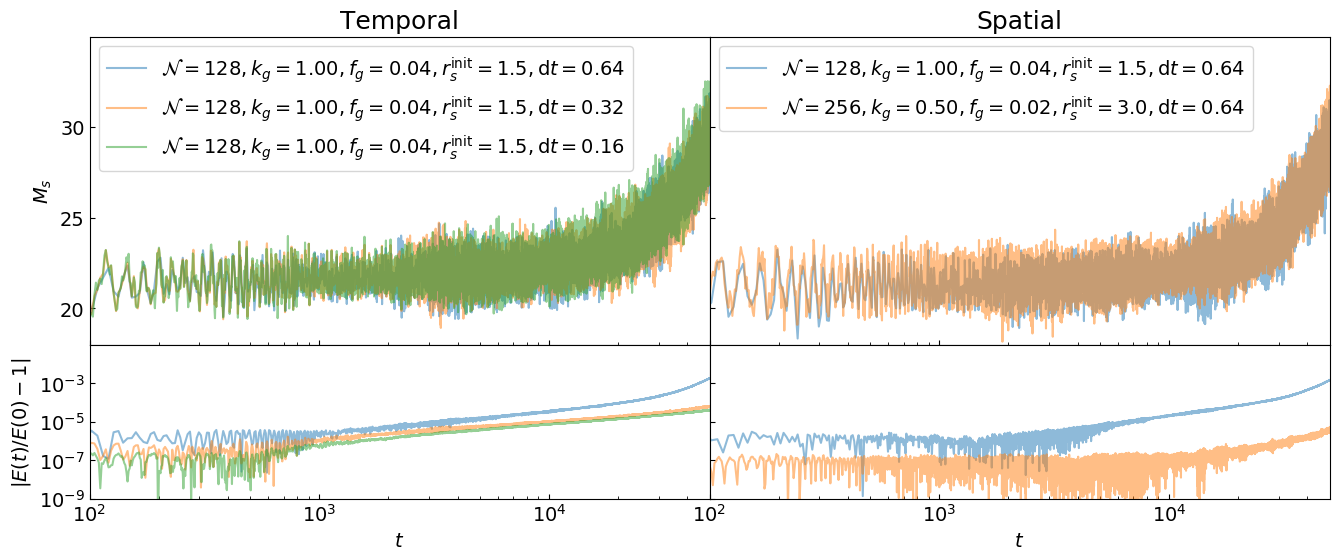}
\caption{Temporal (left) and spatial (right) convergence tests for the
  extreme values of the gas momentum and soliton radius $k_g=1$,
  $\rinit=1.5$. 
Temporal test contains three simulations by decreasing the time step size
$\dt$ by 2 or 4 relative to the fiducial value, 
whereas the spatial test consists of two simulations with  
the box size $\Nres$ differing by a factor of 2. 
The simulations for spatial test follow the scaling
relation (\ref{eq:scaling}). 
\label{fig:resolution}
 }
  \end{center}
\end{figure}

For a third test, we superimpose the soliton and the gas and again run
three simulations with decreasing time step. We take the soliton with
$\rinit=1.5$ and push the gas momentum up to $k_g=1$. The evolution of
the soliton mass and the total energy in these runs is shown in the
left panel of \cref{fig:resolution}. The soliton mass growth in the
three cases is broadly the same, though detailed features are slightly
different. The energy error is low in the initial time range
$t\lesssim 10^3$ where it also obeys the $({\rm d}t)^2$ scaling. However,
from $t\sim 10^3$ it starts to steadily grow and its scaling with
$({\rm d}t)^2$ gets violated. Still, the error remains small until very late
times. For the fiducial time step it reaches $10^{-3}$ when the
soliton mass exceeds $M_s\simeq 27$ and hence its radius drops below
$r_s\simeq 1.2$. This suggests that the soliton-gas system with
$r_s\sim 1.2$ and $k_g\sim 1$ is at the extreme of our numerical
resolution. Since we are interested in the averaged properties of the
soliton evolution, rather than fine details, we accept $k_g=1$ as the
upper boundary for admissible gas momenta. To ensure the absence of
any excessive loss of precision, we monitor the energy conservation
throughout our simulations and only use data where the energy is
conserved with accuracy better than $10^{-3}$.

Finally, we perform a spatial convergence test. 
Instead of varying ${\rm d}x$, which is fixed to $1$ in our
code, we make use of the scaling  
symmetry~(\ref{eq:scaling}). It implies that
decreasing ${\rm d}x$ is equivalent to an increase of ${\cal N}$
accompanied by an appropriate rescaling of other parameters. Thus we
consider two simulation runs with 
$(\Nres=128,\, k_g=1,\, f_g=0.04,\, \rinit=1.5)$ and
$(\Nres=256,\, k_g=0.5,\, f_g=0.02,\, \rinit=3.0)$.
Note that we do not rescale the time step ${\rm d}t=2/\pi$ which
is tied to the lattice spacing in order to avoid aliasing 
\cite{May:2021wwp}. 
The results of these two runs are compared in the right panel of  
\cref{fig:resolution}. While energy conservation is much better
satisfied on the bigger lattice, the broad features of the mass
evolution in these two runs agree. This further support the validity
of our numerical results up to the extreme values $k_g=1$,
$\rinit=1.5$.

\subsection{Conversion of peak density into soliton mass}
\label{sec:peak}

As discussed in \cref{sec:setup}, we estimate the mass of the soliton
and its radius from the maximal density in the box $\rhop$, assuming
that it corresponds to the soliton peak density $\rhoc$. However,
the interference of the soliton wavefunction with the gas waves can
increase the maximal density above that of the soliton. The
increase is proportional to the product of the soliton and gas 
wavefunctions, hence to
the geometric mean of their densities. In more detail, we can
estimate the bias as 
\be
\frac{\rhop}{\rhoc}-1\sim 2\sqrt{\frac{\rho_g}{\rhoc}}\;,
\ee
which can be significant even for large density contrasts. For
example, the density bias is about $40\%$ for
$\rhoc/\rho_g=30$. The situation is further complicated by large
fluctuations in the local gas density that can further increase the
bias. In particular, when the soliton is too light, its peak becomes
completely obscured by the gas. 
 
\begin{figure}[t]
\begin{center}
 \includegraphics[width=0.75\textwidth]{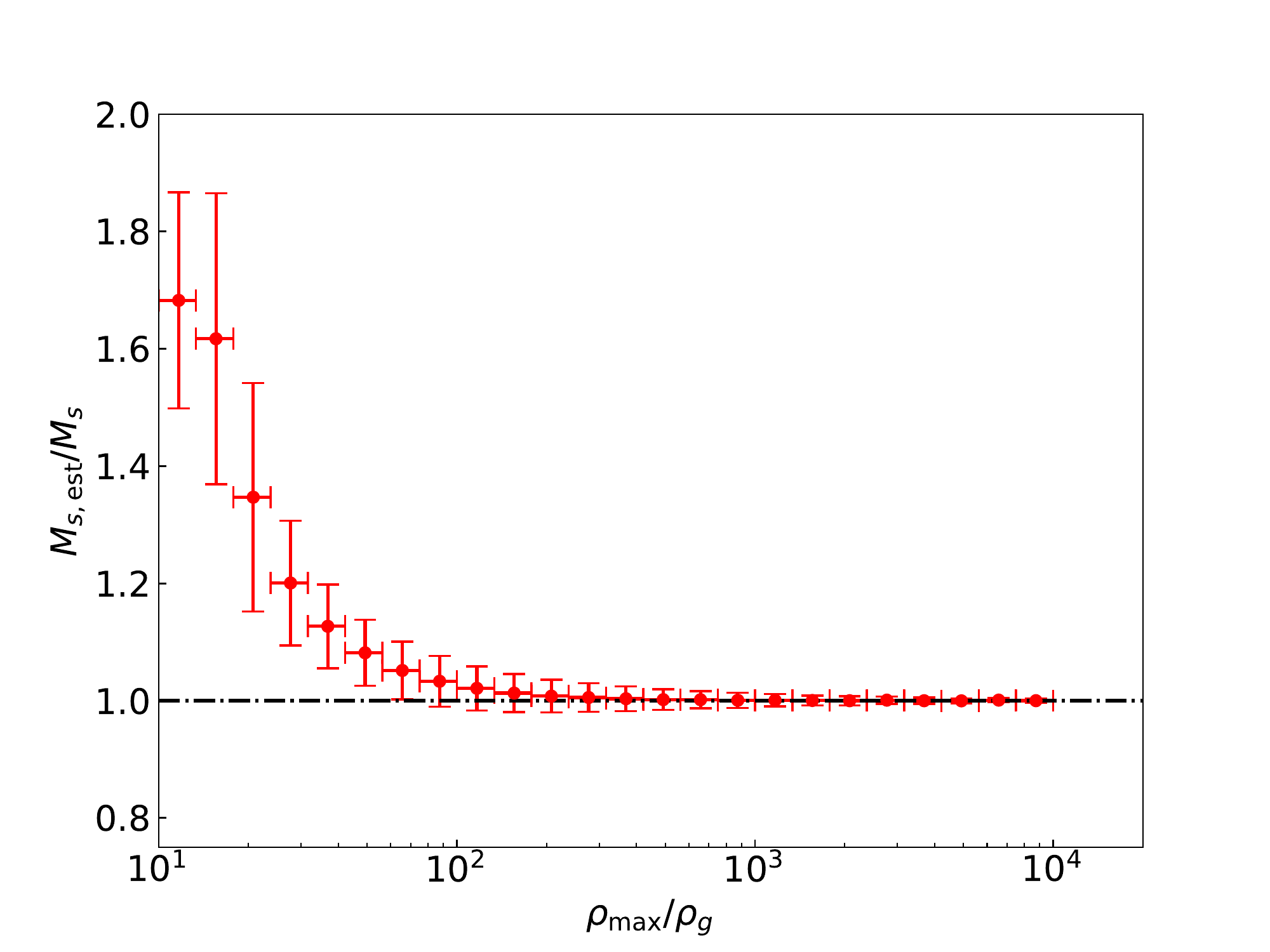}
\caption{Ratio of the soliton mass estimator to the true soliton mass
  as functon of the density contrast in the axion field generated by
  superposition of the soliton and gas wavefunctions. We adopt the threshold 
$\rhop>30\,\rho_g$ when measuring the soliton mass from the
simulations.
\label{fig:peak2mass}
}
  \end{center}
\end{figure}

To pin down the lowest density contrast between the soliton and the
gas for which the bias is unimportant, we conduct a series of the
following auxiliary numerical experiments. We generate a gas field
with given mean density $\rho_g$ and superimpose on it a soliton of
mass $M_s$ {\em without any evolution}. Then we evaluate the estimator
of the soliton mass using our formula
\begin{equation}
  \Mest = 25.04 \, \rhop^{1/4}\;,
  \label{eq:rhos2mc}
\end{equation}
where $\rhop$ is the maximal density of the axion field in the
box. The estimator is compared to the true soliton mass in 
\cref{fig:peak2mass}.
We observe that when the soliton is prominent enough, say 
 $\rhoc\simeq\rhop>100\,\rho_g$, the estimator is
 unbiased.
On the other hand, for 
$\rhop\lesssim20\,\rho_g$, we are essentially unable to distinguish the
soliton peak against the gas density fluctuations. We adopt the
threshold 
$\rhop>30\,\rho_g$ when measuring the soliton mass in our simulations, which 
introduces an error of at most 20\% in the mass estimate.

%==========================================================================
\bibliography{axion}
\bibliographystyle{JHEP}

%==========================================================================

\end{document}